\documentclass[aps, prd,10pt,notitlepage,nofootinbib,superscriptaddress,showkeys,groupedaddress]{revtex4-1}
\usepackage{amstext,amsmath,amssymb,amsfonts,bbm}
\usepackage[latin1]{inputenc}
\usepackage{graphicx}
\usepackage{hyperref}
\usepackage{epstopdf}
\usepackage{amsthm}
\usepackage{pstricks}
\usepackage{tocvsec2}

\usepackage{color}

\def\beq{\begin{equation}}
\def\be{\begin{equation}}
\def\ee{\end{equation}}
\def\bes{\begin{eqnarray}}
\def\ees{\end{eqnarray}}

\DeclareMathOperator{\Inv}{Inv}

\DeclareMathOperator{\im}{im}
\DeclareMathOperator{\Ad}{Ad}
\DeclareMathOperator{\Hom}{Hom}
\DeclareMathOperator{\rk}{rk}
\DeclareMathOperator{\id}{id}
\DeclareMathOperator{\SU}{SU}
\DeclareMathOperator{\vol}{vol}
\DeclareMathOperator{\U}{U}
\DeclareMathOperator{\tor}{tor}
\DeclareMathOperator{\Vol}{Vol}

\newcommand{\alg}{\mathfrak{g}}

\renewcommand{\u}{\mathfrak{u}}

\newcommand{\unit}{\mathbbm{1}}

\def\bra{\langle}
\def\ket{\rangle}
\def\f{\frac}

\def\mone{^{-1}}

\def\pp{\partial}

\def\C{{\mathbbm C}}

\def\calA{{\mathcal A}}

\def\calF{{\mathcal F}}

\def\calH{{\mathcal H}}

\def\calO{{\mathcal O}}

\def\calZ{{\mathcal Z}}

\def\u{\underset}
\def\k{K_{\tau}}

\theoremstyle{definition}\newtheorem{definition}{Definition}
\theoremstyle{definition}\newtheorem{theorem}{Theorem}
\theoremstyle{definition}
\theoremstyle{definition}
\theoremstyle{definition}
\theoremstyle{definition}

\begin{document}
\maxtocdepth{subsection}

\title{\large \bf Bubble divergences from twisted cohomology}

\author{{\bf Valentin Bonzom}}\email{valentin.bonzom@ens-lyon.fr}\author{{\bf Matteo Smerlak}}\email{smerlak@cpt.univ-mrs.fr}

\affiliation{Centre de Physique Th\'eorique\\ Campus de Luminy, Case 907, 13288 Marseille Cedex 09 France}

\date{\small\today}

\begin{abstract}\noindent
We consider a class of lattice topological field theories, among which are the weak-coupling limit of 2d Yang-Mills theory and 3d Riemannian quantum gravity, whose dynamical variables are flat discrete connections with compact structure group on a cell 2-complex. In these models, it is known that the path integral measure is ill-defined because of a phenomenon known as `bubble divergences'. In this paper, we extend recent results of the authors to the cases where these divergences cannot be understood in terms of cellular cohomology. We introduce in its place the relevant {\it twisted} cohomology, and use it to compute the divergence degree of the partition function. We also relate its dominant part to the Reidemeister torsion of the complex, thereby generalizing previous results of Barrett and Naish-Guzman. The main limitation to our approach is the presence of singularities in the representation variety of the fundamental group of the complex; we illustrate this issue in the well-known case of two-dimensional manifolds.
\end{abstract}

\maketitle
\tableofcontents


\section{Introduction}

One road to the quantization of a background-independent field theory such as general relativity is the {\it spinfoam formalism}. In this approach, which can be thought of as the covariant formulation of loop quantum gravity \cite{rovelli-book,thiemann-book}, the Feynman path integral is realized as a sum of amplitudes associated to oriented two-dimensional cell complexes, aka \emph{foams}. It is expected that this non-perturbative expansion could be free from the inconsistencies of perturbative quantum gravity (although outdated, \cite{baez-bf-spinfoam,perez-spin-foam-representation} remain good reviews; see also \cite{fk-action-principle}).

Inspired by the Plebanski's observation \cite{plebanski} that general relativity can be interpreted as a constrained BF theory with a simple\footnote{A $2$-form $B$ is {\it simple} if can be written $B=e\wedge e$ for a certain $1$-form $e$.} $B$-field, the construction of spinfoam models for four-dimensional quantum gravity has followed a remarkable pattern \cite{epr-long,new-model-fk}. One starts from a `flat' spinfoam model analogous the one proposed by Ponzano and Regge \cite{PR}, with {\it flat} gauge connections with a compact structure group $G$ as dynamical variables; Fourier analysis on the group then yields a formulation of the amplitudes as discrete sums over quantum numbers labelling the edges ($1$-cells) and faces ($2$-cells) of cell $2$-complexes; to obtain an ansatz for a quantum gravitational amplitude, one eventually imposes certain `simplicity' constraints \`a la Plebanski restricting the range of these quantum numbers. 

In \cite{perez-BC-bubble}, however, it was pointed out that the amplitudes of the flat spinfoam model are not well-defined, because of a phenomenon coined {\it bubble divergences}, and related to the presence of certain sets of faces in the $2$-complex forming {\it closed} surfaces. This interpretation was then strengthened in \cite{freidel-louapre-diffeo}, where the bubble divergences were related to discrete Bianchi identities. With the understanding that these bubble divergences need a renormalization procedure, partial powercounting theorems were proved in \cite{freidel-gurau-oriti,ben-geloun-abelien}, based on a combinatorial counting of these `bubbles'. 

In our previous paper \cite{cell-homology}, we have discussed in more detail the accuracy of the intuition that the divergence degree of a foam $\Gamma$ can be inferred from its combinatorial, or topological structure. Defining bubbles as $2$-cycles in the cellular cohomology of $\Gamma$, we have showed that there are cases where the divergence degree of $\Gamma$ is indeed given by its second cellular Betti number (its `number of bubbles'): when $G$ is Abelian, or when $\Gamma$ is simply connected. But, more importantly, we have also explained why such ideas are deceiving in more general situations. In short, the amplitudes can be reduced to integrals over the space of {\it flat} $G$-connections on $\Gamma$, which is also the representation variety of its fundamental group in $G$, and the structure of this space involves {\it both} the topology of $\Gamma$ and the non-Abelian structure of $G$ in a very non-trivial way. 

Here, we provide a finer description of these `bubble divergences', using the tools of {\it twisted cohomology}. In particular, when the singularities of the representation variety of $\pi_{1}(\Gamma)$ can be neglected (in a sense explained in sec. \ref{divergence}), we find that the divergence degree of a foam is given by the value of the second twisted Betti number on generic flat connections. This result allows to extract a dominant part from the amplitudes, which we relate to the Reidemeister torsion of $\Gamma$. To illustrate these ideas, we use the weak-coupling limit of 2d Yang-Mills theory -- a particular case of the flat spinfoam model, where these ideas were explored by Witten \cite{witten-2d-YM}, Goldman \cite{goldman} and many others. 

On the mathematical side, the flat spinfoam model has proved to be connected to a number of interesting problems in algebraic topology. In addition to the two-dimensional case, where it provides a very efficient way to compute the volume of the moduli space of flat connections \cite{witten-2d-YM}, remarkable topological invariants have been obtained in three and four dimensions on its basis. On $3$-manifolds, the Turaev-Viro invariant
\cite{turaev-viro} can be seen as a regularization of the Ponzano-Regge model by a cosmological constant. In four dimensions, the Crane-Yetter invariant  \cite{crane-kauffman-yetter} provides a combinatorial way to compute the signature
of the manifold. Note however that in these two examples, the compact Lie group of the original flat spinfoam model is replaced by a quantum group. Our analysis in this paper can be understood as a way to properly define the flat model {\it without} quantum groups, for a general cell 2-complexes (as opposed to the $2$-skeleton of the dual cell complex to a triangulated manifold, as in the Turaev-Viro and Crane-Yetter models).

Yet another motivation to study the bubble divergences of the flat spinfoam model is the following.
As observed by Boulatov \cite{boulatov-model} and Ooguri
\cite{ooguri4d}, its amplitudes can be interpreted
as Feynman amplitudes of a certain non-local field theory defined over
a Cartesian power of the structure group, known as {\it group field
theory}. The study of group field theory is considered a promising
approach to four-dimensional quantum gravity
\cite{freidel-gft,oriti-gft-review-06}. One virtue of this approach is
that by summing over Feynman `graphs', which are really cell 2-complexes, the
topology of spacetime is not a priori fixed, but could emerge dynamically. From this perspective, bubble divergences appear as a higher categorical version of the usual ultraviolet divergences, and their renormalization, attempted in \cite{freidel-gurau-oriti,ben-geloun-abelien}, can be properly understood in the usual field-theoretic sense.

The paper is organized as follows. In sec. \ref{connections}, we introduce in detail the flat spinfoam model, and the relevant tools from twisted cohomology. We use them in sec. \ref{divergence} to compute the divergence degree of a foam away from singularities, and illustrate how the latter can be dealt with in the two-dimensional case. In sec. \ref{reidemeister}, we relate the dominant part of the amplitudes to the Reidemeister torsion. Our conclusion follows in sec. \ref{sec:conclusion}.

\section{Discrete connections and twisted cohomology}\label{connections}

In this introductory section, we define the flat spinfoam model, review the current understanding of its `bubble divergences', and introduce the twisted cohomology induced on the foams by discrete flat connections.
\subsection{The flat spinfoam model}

Let us present in more details the generalization of the Ponzano-Regge \cite{PR} or Ooguri \cite{ooguri4d} models which we refer to as the `flat spinfoam model'. In both latter cases, one starts with a triangulated (spacetime) manifold (of dimension 3 and 4, respectively), and uses the 2-skeleton of the dual cell complex to assign it a quantum amplitude. Both from the perspective of loop quantum gravity and of group field theory, however, it is natural to consider amplitudes defined on \emph{arbitrary} oriented cell 2-complexes, whether dual to manifolds or not. We call \emph{foams} such complexes. In this paper, we will only consider foams without boundary, which would arise in the Ponzano-Regge and Ooguri models in the case of closed manifolds. Let $\Gamma$ be such a closed foam. We denote $\Gamma_i$ ($i=0,1,2$) the set of its $i$-cells (vertices, edges and faces respectively), and $V=\vert\Gamma_0\vert$, $E=\vert\Gamma_1\vert$, $F=\vert\Gamma_2\vert$. We also need a structure group $G$, which we take as a compact semi-simple Lie group with Lie algebra $\alg$. 

A \emph{connection on the foam} $\Gamma$ is the assignment of an element $g_e$ of the structure group $G$ to each edge of $\Gamma$.  These elements can be thought of as holonomies or `parallel transport' operators between vertices of the foam. This is also the standard way to discretize a genuine connection on a cell decomposition of a manifold. The space of connections on $\Gamma$ is therefore
\be
\calA \,\equiv\, \bigl\{ A = (g_e)_{e\in\Gamma_1}\,\in G^E\bigr\},
\ee
The curvature of a connection $A$ is encoded in the holonomies along the boundary of faces. It is represented by the map 
\beq \label{curvature map}
\begin{aligned}
H\ :\ &\calA\rightarrow G^F\\
&A\mapsto  \Bigl( H_f(A)=\prod_{e\in\partial f} g_e^{[f:e]} \Bigr)_{f\in\Gamma_2},
\end{aligned}
\ee
where $[f:e]$ is the incidence number of the face $f$ on the edge $e$, and $H_f(A)$ is the `holonomy' of the connection $A$ around the face $f$. This provides a notion of flatness on the foam: the connection is \emph{flat} if\footnote{Throughout this paper, $\unit$ denotes the unit element of the relevant group.}
\be \label{flatness}
H(A)\,=\,\unit.
\ee

The {\it flat spinfoam model} is then defined formally as the partition function of a system of flat $G$-connections on $\Gamma$:
\be \label{amplitude}
\calZ(\Gamma,G) \,=\, \int_{\mathcal{A}} dA\prod_{f\in\Gamma_2} \delta\bigl(H_f(A)\bigr),
\ee
where $dA = \prod_{e\in\Gamma_1} dg_e$ is the Haar measure on $\mathcal{A}=G^E$, and $\delta(g)$ is the Dirac delta on $G$. Obviously, the support of this integral is the set of flat connections
\be
\calF \,\equiv\, H^{-1}(\unit).
\ee
Since the curvature map $H$ is smooth, $\calF$ is a smooth manifold of $\calA$ whenever the unit $\unit$ is a regular value of $H$. But this is generally not the case, as we will see.

The spinfoam formalism consists in rewriting the partition function of the model defined in \eqref{amplitude} with integrals as sums over data labelling the cells of $\Gamma$ and coming from the representation theory of $G$. Obviously, it is likely that the product of delta ditributions in \eqref{amplitude} is not well-defined (and to study this issue is the goal of the present article), but let us just ignore this difficulty for a few paragraphs. The spinfoam way to compute $\calZ(\Gamma,G)$ starts with the spectral decomposition of the Dirac delta over $G$
\beq
\delta(g) = \sum_\rho \dim(\rho)\,\chi_\rho(g),
\ee
where the sum runs over (equivalence classes of) irreducible representations $\rho$ of $G$ with characters $\chi_\rho$. This assigns representations $\rho_f$ to every dual face, so that
\beq \label{expdirac}
\calZ(\Gamma,G) = \sum_{(\rho_f)_{f\in\Gamma_2}}\,\int_{\cal A} dA\ \prod_{f\in\Gamma_2} \dim(\rho_f)\ \chi_{\rho_f}\bigl(H_f(A)\bigr).
\ee
Then, for a fixed set of representations, one can perform the integrals over the $E$ copies of the group, by tensoring the representations and using their orthogonality. 

To see how this goes, let us assume now that $\Gamma$ is the 2-skeleton of a cell decomposition dual to a triangulated $d$-dimensional manifold $M$.
Then, faces of $\Gamma$ are dual to $(d-2)$-simplices, while edges of $\Gamma$ are dual to $(d-1)$-simplices. Since the latter have $d$ $(d-2)$-simplices on their boundary, each edge of $\Gamma$ is shared by $d$ faces. Thus, after expanding the characters on matrix elements of the group elements $(g_e)$, one has to integrate over exactly $d$ matrix elements for each of them. In two dimensions, the Schur orthogonality relation implies that all representations in \eqref{expdirac} are equal to, say, $\rho$. Introducing the Euler characteristic of the surface, $\chi(M) = F-E+V$, the partition function then reads
\beq \label{sf 2d}
\calZ(\Gamma,G) \,=\, \sum_\rho\dim(\rho)^{\chi(M)},
\ee
which is obviously independent of the chosen triangulation of $M$. This formula was shown by Witten \cite{witten-2d-YM} to be the most efficient way to compute the symplectic volume of the moduli space of flat connections for closed orientable surfaces (see sec. \ref{YM}). Also, for a given group, it is easy to see whether it gives a finite answer or not (in particular, it is always divergent for the 2-sphere and the 2-torus).

Things go differently in higher dimensions. Denoting $\calH_\rho$ the carrying space of the representation $\rho$, the integral of $d$ matrix elements over $G$ is now given by
\beq
\int_G dg\ D^{(\rho_1)}(g)\otimes \dotsm\otimes D^{(\rho_d)}(g) = P_{\Inv(\calH_{\rho_1}\otimes\dotsm\otimes\calH_{\rho_d})}.
\ee
Here the right hand side is the orthogonal projector on the space $\Inv(\calH_{\rho_1}\otimes\dotsm\otimes\calH_{\rho_d})$ of invariant tensors (intertwiners) in the product of the representations $\rho_f$ meeting at an edge $e$. This projector can be expanded onto a basis of intertwiners $\iota_e : \otimes_{f=1}^d \calH_{\rho_f}\rightarrow \C$, so that edges are finally labelled by intertwiners between the $d$ representations meeting at each of them. For instance, if $M$ is a three-dimensional manifold and $G=\SU(2)$, the irreducible representations on the faces are labelled by spins $(j_f)_{f\in\Gamma_2}$, and all the edges are colored by the only invariant in $\calH_{j_1}\otimes\calH_{j_2}\otimes\calH_{j_3}$ up to normalization, the Wigner $3mj$-symbol. Intertwiners are contracted with each one another among the $4$ edges meeting at each vertex of $\Gamma$. This gives rise to a so called `vertex amplitude', which by duality assigns an amplitude to any tetrahedron. This yields the Ponzano-Regge model \cite{PR}
\beq \label{pr model}
\calZ_{\rm PR}(\Gamma,\SU(2)) = \sum_{(j_f)_{f\in\Gamma_2}} \prod_f (-1)^{2j_f} \bigl( 2j_f+1\bigr) \prod_v W_v^{\rm PR}(j_f),
\ee
for which the vertex amplitude is the Wigner $6j$-symbol,
\beq
W_{v}^{\rm PR}(j_f) = (-1)^{\sum_{i=1}^6 j_i}\ \begin{Bmatrix} j_1 &j_2 &j_3\\ j_4 &j_5 &j_6 \end{Bmatrix}.
\ee
The Ponzano-Regge model is often interpreted as a state-sum model for Riemannian 3d quantum gravity in the Palatini-Cartan formulation, i.e. with possibly degenerate metrics. Since the latter is nothing but topological BF field theory, one expects the partition function $\calZ_{\rm PR}(\Gamma,\SU(2))$ to be independent of the triangulation, and thus to define a $3$-manifold invariant. However, this expectation was never made precise, because the sums over spins are usually divergent.

To summarize, if the spinfoam formalism gives in two dimensions a nice way to compute \eqref{amplitude}, which is the volume of the space of flat connections \eqref{sf 2d}, it is more difficult to obtain analogous results in highers dimensions. It turns out that, in these cases, the initial expression \eqref{amplitude} will be more useful to study this aspect and understand the structures of divergences.

\subsection{Foams and bubbles}\label{foams and bubbles}


Typical bubble divergences in the flat spin foam model arise when trying to prove topological invariance, i.e. invariance under changes of the triangulation for a given manifold. For instance, in three dimensions, there are two elementary moves (together with their inverses) which enable to relate triangulations of homeomorphic manifolds: the 2-3 Pachner move, which transforms two adjacent tetrahedra into three, and the 1-4 Pachner move, which transforms a single tetrahedron into four. More precisely, the 1-4 move consists in adding a vertex in the bulk of a tetrahedron, and relate it to the four other vertices, thus creating four new edges in the triangulation. In the dual picture, this process gives a 3-cell dual to the added inner vertex, whose boundary is made of four triangular faces -- pictorially a \emph{bubble}. Quite clearly, these four faces are not independent, in the sense that imposing the triviality of the holonomy $H_f$ around any three of them automatically enforces triviality for the remaining face. Thus, taking the formula \eqref{amplitude} literally, this gives a divergent factor $\delta(\unit)$. In the spinfoam formalism, this factor arises as $ \sum_{\rho} (\dim\rho)^2$. In fact, this phenomenon is to be expected for all inner vertices of the triangulation. Indeed, if $F_v$ is the set of dual faces of $\Gamma$ `wrapping around' an inner vertex $v$ of the triangulation, then it is easy to convince oneself to there will be an ordering of $F_v$ such that
\be \label{bubble group variables}
\prod_{f\in F_v} H_f^{\epsilon_f} = \unit,
\ee
where $\epsilon_f$ is $\pm1$. In short, a holonomy around a face of $F_v$ can be expressed as a product of the others. This can be seen as a discrete version of the Bianchi identity for the curvature of a connection. This heuristic reasoning led Freidel and Louapre \cite{freidel-louapre-diffeo} to a simple process designed at regularizing this type of divergences. The idea is simply to remove the group Dirac deltas along a spanning tree of the triangulation (that is a set of edges touching every vertex without forming a loop).

However, an example detailed in \cite{barrett-PR} shows that this process may not be sufficient: the Ponzano-Regge state-sum on Bing's house with two rooms is not absolutely convergent even though there are no inner vertices in this triangulation of the $3$-ball.\footnote{Note this example falls into the case of simply connected complexes treated in \cite{cell-homology}.} Furthermore, this type of divergence is specific to the 3d case: its extension to higher dimensional manifolds is somewhat subtler. Indeed, its natural extension in dimension $d$ would be that every $(d-3)$-simplex, being dual to a 3-cell, would contribute a divergent factor $\delta(\unit)$. But this is already false in dimension 4, as one can observe when applying a 1-5 Pachner move on a single 4-simplex. This consists in adding a vertex to the triangulation within a 4-simplex, and to link it to the five other vertices of the simplex. This generates five 3-cells, while there are only four redundant deltas \cite{ooguri4d}. This is because the five 3-cells are not independent: they form the boundary of a 4-cell, dual to the vertex added to the triangulation. This phenomenon is nothing but a discrete analog of the reducibility of the gauge symmetry of the topological BF field theory \cite{blau-thompson-bf}. This interesting view of divergences on manifolds will be studied more precisely elsewhere.

Finally, we have argued that the natural setting to define the flat spin foam model uses cell 2-complexes, where the notion of 3-cells (and higher-dimensional cells) is obviously meaningless. Thus, we need a definition of a bubble which catches the intuitive picture of a 3-ball, using only two-dimensional objects. This means looking at some relations between faces. Recently, a first definition of a bubble has been proposed by Gurau \cite{gurau-colored-gft} in a purely graphical way, which leads him to discuss the divergence degrees of group field theory `graphs' in terms of `bubble homology'. We found it more powerful in \cite{cell-homology}, instead of that graphical method, to use a standard algebraic notion in topology: the cellular homology of the 2-complex, and consequently to propose the following
\begin{definition}
A {\it bubble} is a $2$-cycle in the cellular homology of a foam.
\end{definition}

With this definition, the `number of bubbles' is the \emph{second Betti number} $b_2(\Gamma)$, that is the dimension of the second homology group $H_2(\Gamma)$. Describing typical situations where the divergence degree is exactly given by this number of bubbles was the purpose of \cite{cell-homology}. But it is easy to see that, unfortunately, only a limited number of situations can be controlled with cellular homology. Let us consider again the two-dimensional case, for the torus. Choosing the standard cell decomposition with two edges (the non-trivial cycles) and one face, we have
\beq
\calZ(S^1\times S^1,G) = \int_{G^2} da\,db\ \delta([a,b]),
\ee
where the group commutator is $[a, b] \equiv aba\mone b\mone$. Clearly, the Abelian case $G=\U(1)$ is special because it trivializes the commutators, leading to a formal divergent factor $\delta(\unit)$. The simplest non-Abelian case, $G=\SU(2)$, exhibits a different and interesting behaviour. The condition $[a, b]=\unit$ constrains only two parameters among the three of each group elements: their class angle remains free. Indeed, writing $g=\exp(i\psi_g\,\hat{n}_g\cdot\vec{\sigma})\in\SU(2)$, where $\vec{\sigma}$ is the vector made of the three Pauli matrices\footnote{The Pauli matrices are Hermitian matrices, satisfying the commutation law $[\sigma_i,\sigma_j]=2i\epsilon_{ij}^{\phantom{ij}k}\,\sigma_k$, for $i,j=1,2,3$, with $\epsilon_{ij}^{\phantom{ij}k}$ the completely antisymmetric tensor.} and $\hat{n}_g\in S^2$ is the direction of the rotation,
\beq
[\exp\bigl(i\psi_a\,\hat{n}_a\cdot\vec{\sigma}\bigr), \exp\bigl(i\psi_b\,\hat{n}_b\cdot\vec{\sigma}\bigr) ] =\unit\quad \Leftrightarrow\quad \hat{n}_a =\pm\hat{n}_b.
\ee
Thus, the three-dimensional delta over $\SU(2)$ can only be used to integrate, say, the direction of $\hat{n}_b$, which leaves a component of the constraint trivially satisfied. Note that this explains the divergence of \eqref{sf 2d} in this situation. Moreover, this divergence is clearly not of the type $\sum_j (2j+1)^2$, but rather $\sum_j (2j+1)^0$.

In comparison with the discussion at the beginning of the section, this shows that one has to carefully examine the linear relations which may exist between the \emph{real} components of the conditions $H_f(A)=\unit$. Also, it shows that divergences cannot be extracted by just looking at the cell complex: they involve some non-trivial interaction between the foam and the structure group. This will lead us to introduce a \emph{twisted homology}, already known to be relevant for 2d Yang-Mills theory and more recently for the Ponzano-Regge model \cite{barrett-PR}.

\subsection{Gauge transformations}

We saw that a discrete connection $A\in\mathcal{A}$ can be viewed as a collection of $E$ group elements, $A=(g_e)_{e\in\Gamma_1}$, representing the parallel transport operators along the edges of $\Gamma$. A discrete gauge transformation $h$ is therefore a set of $V$ group elements $(h_v)_{v\in\Gamma_0}$ acting at the vertices of $\Gamma$ according to
\be
h\,\cdot\, A = \bigl(h_{t(e)}\,g_e\,h^{-1}_{s(e)}\bigr)_{e\in\Gamma_1}.
\ee
When $\Gamma$ is the 2-skeleton of the dual cell complex to a triangulation of a manifold, this is indeed the effect of a gauge transformation on the parallel transport operators of a genuine connection in the continuum. 

We will be interested in `factoring out' gauge transformations to identify equivalent flat connections: those lying along the same orbit of the $G^V$-action. As a first step, it is convenient to reduce gauge transformations so that they act at \emph{one} vertex only. This can be done through the standard process of contracting a maximal tree of $\Gamma$, that is a tree touching every vertex of $\Gamma$ whitout forming loops.\footnote{This will not affect the divergence degree of the foam, which turns out to be homotopy invariant, see below.}

\begin{definition}
If $\Gamma$ is a foam with $V$ vertices, $E$ edges and $F$ faces, its \emph{reduction} is the deformation retract of $\Gamma$ with $1$ vertex, $E-V+1$ edges and $F$ faces. 
\end{definition}

In the following we will only consider reduced foams $\Gamma$ of this kind, hence with $V=1$, except when explicitly stated. A gauge transformation is then just the conjugation of the elements $g_e$ by a single element $h\in G$:
\be
h\,\cdot\,A = \bigl(h\, g_e\, h\mone\bigr)_{e\in\Gamma_1}.
\ee
To evaluate the effect of a \emph{small} gauge transformation, we consider the differential at $h=\unit$ of the map
\beq \label{gauge map}
\begin{aligned}
\gamma_A\ :\ &G\ \rightarrow G^E \\
&h\ \mapsto\ \bigl(h\,g_e\,h^{-1}\bigr)_{e\in\Gamma_1}.
\end{aligned}
\ee
It is given by
\beq
\begin{aligned}
d\gamma_{A|\unit}\ :\ &\alg\ \rightarrow T_A\, G^E \\
&v\ \mapsto\ \Bigl(R_{g_e*}\bigl(\id-\Ad_{g_e}\bigr)v\Bigr)_{e\in\Gamma_1},
\end{aligned}
\ee
where $\Ad$ stands for the adjoint representation of $G$ on $\alg$, and $R_{g*}$ for the right translation. In matrix notation: $\Ad_g v = gvg^{-1}$ and $R_{g*}v = vg$.

The kernel of $d\gamma_{A|\unit}$ is the algebra of the isotropy group $\zeta(A)$ of the connection $A$, while its image corresponds to the directions along which $A$ is changed by the group action. If $\calO_A$ denotes the orbit through $A$, then we know that
\be
\calO_A \simeq G/\zeta(A),\qquad\text{and}\qquad T_A\,\calO_A = \im d\gamma_{A|\unit}.
\ee
The following cases can be distinguished:
\begin{definition}
A connection $A$ is said to be {\it reducible} if it admits a non-trivial isotropy group, or {\it irreducible} if it is only preserved by the center $\zeta(G)$ of $G$.
\end{definition}
So an irreducible connection is characterized by the fact that $d\gamma_{A|\unit}$ is of maximal rank, $\rk d\gamma_{A|\unit} =\dim G$, and the orbit through it is $\calO_A \simeq G/\zeta(G)$. For $G=\SU(2)$, the reducible connections are precisely the \emph{Abelian} connections, such that the set $(g_e)_{e\in\Gamma_1}$ lives in a subgroup $T\simeq \U(1)$ of $\SU(2)$, or equivalently $\rk d\gamma_{A|\unit} =2$. In this case, the orbit is the homogeneous space $S^2=G/T$. Finally, trivial connections, such that $A\in \zeta(G)^E$, are called \emph{central} connections. They are left invariant by $G$ itself, and the rank of $d\gamma_{A|\unit}$ is zero.

\subsection{Flat connections and twisted cohomology} \label{sec:twistedhom}

Since our amplitudes are supported only on the set of flat connections $\cal F$, let us now further our insight into the structure of this space. One useful way to describe it is in terms of the fundamental group $\pi_1(\Gamma)$ of the 2-complex. Indeed, given a cell 2-complex $\Gamma$, one can find a presentation of its fundamental group by retracting a spanning tree, as described above: then, the generators $a_{e}$ of this group correspond to the edges $e$, and there is one relation per face which is formally exactly the flatness condition \eqref{flatness} for the generators of $\pi_1(\Gamma)$:
\be
\pi_{1}(\Gamma)=\bra(a_{e})_{e\in\Gamma_{1}}\vert(\prod_{e}a_{e}^{[f:e]})_{f\in\Gamma_{2}}=\unit \ket.
\ee
Notice that the relationship between presentations of groups and 2-complexes goes both ways: a finite presentation of a group $\pi$ unambiguously determines a complex $\Gamma$. From a single vertex, draw an edge for each generator, and attach the faces according to the relators.\footnote{Note that trivial relations such as $aa^{-1}=\unit$ must not be eliminated from the presentation of $\pi$ for this duality to hold. An example of this issue is provided by the `dunce hat': while $\bra a\vert a^2a^{-1}=\unit\ket$ is obviously equivalent to $\bra a\vert a=\unit\ket$ as a group presentation, the corresponding $2$-complexes, the dunce hat and the disc respectively, are not.} In the following we will use this process to identify a cell-complex $\Gamma$ with one vertex and a presentation of its fundamental group $\pi_1(\Gamma)$. The natural question is then to understand to which extend the amplitude $\calZ(\Gamma,G)$ does depend on the chosen presentation of $\pi_1(\Gamma)$. We will examine this question in section \ref{sec:invariance}, and show that the dominant part of $\calZ(\Gamma,G)$, defined below, is invariant under certain changes on the presentation corresponding to 2-deformations of the foam $\Gamma$. Later on, in section \ref{reidemeister}, we will see that the dominant part of $\calZ(\Gamma,G)$ can be furthermore expressed, under some assumptions, in terms of the Reidemeister torsion of $\Gamma$, which is known to be an invariant of simple-homotopy.


The previous argument shows that a flat connection on $\Gamma$ can be seen as a homomorphism from $\pi_1(\Gamma)$ to $G$,
\be \label{representation space}
\calF\simeq \Hom\bigl(\pi_1(\Gamma),G\bigr).
\ee
This space is usually called the \emph{representation variety} of $\pi_1(\Gamma)$ into $G$ (or the space of flat $G$-bundles over $M$, when $\pi_1(\Gamma)$ is seen as the fundamental group of a manifold $M$). It is independent of the chosen presentation of $\pi_1(\Gamma)$. In cases of interest, $G$ is a classical matrix group so that $\calF$ is a \emph{real algebraic set}.

Since the integrand of \eqref{amplitude} is gauge invariant, it is natural to also introduce the moduli space of flat connections on $\Gamma$, i.e. $\calF/G$, also known as the \emph{character variety} of $\pi_1(\Gamma)$, and which is also a real algebraic set. We can distinguish the following situations:
\begin{itemize}
 \item If $\calF/G$ consists in finitely many points, then it means that the tangent space to $\calF$ at a given point is just the tangent space to the orbit of this point. For instance, this happens when $\pi_1(\Gamma)$ is a finite group. This situation is very similar to some cases studied in \cite{cell-homology}, except for some non-Abelian features. However, it can be treated following the same lines \cite{cell-homology}, simply changing all quantites coming from the cellular cohomology with their equivalent in the twisted cohomology we are about to describe.
 \item We will be mainly interested in the case when $\calF/G$ is of dimension at least one. Generically, $\calF$ has several irreducible components. In the following, we will concentrate our analysis on a single component, and keep in mind that in the end one has to sum the contributions from the different irreducible components. So we assume without loss of generality that $\calF$ is an algebraic variety, i.e. that it is irreducible. 
\end{itemize}

To perform the integrals defining our amplitude \eqref{amplitude}, we also need to know the \emph{local} structure of the set of flat connections $\calF$. In particular, the following observation is key to our analysis: $\calF=H^{-1}(\unit)$ decomposes into a smooth submanifod $\calF_0$ of the same dimension as $\calF$ (the `generic' connections), and a set of {\it singular} connections, which is of smaller dimension. The set $\calF_{0}$ can be identified locally using the differential of the curvature map $H$ with the following
\begin{definition} \label{singular}
 A flat connection $\phi$ is \emph{non-singular} if $\dim\ker dH_{\phi}=\dim\calF$. Otherwise, it is {\it singular}, and we have $\dim \ker dH_\phi > \dim T_\phi\calF$.
\end{definition}
The local structure of the space $\calF$ was described by Goldman \cite{goldman}, using the notion of the Zariski tangent space. For what concerns us, it is enough to recall that if $\phi$ is non-singular, then
\be \label{flat tangent space}
T_\phi \calF_{0} = \ker dH_{\phi},
\ee
as one would expect from differential geometry. Let us emphasize however that (\ref{flat tangent space}) does \emph{not} hold if $\phi$ is singular. Since the starting point of our analysis is precisely this relation, the presence of singularities in $\calF$ is the main limitation to our approach. This said, the behaviour of $dH_\phi$ in the neighbourhood of a singular point is an open question for an arbitrary group $\pi_1(\Gamma)$. When $\pi_1(\Gamma)$ is the fundamental group of a closed orientable surface, which is the most studied situation, the variety of representations into $G$ has singularities, but it turns out that they do not affect the partition function. In any similar cases, our method produces the exact divergence degree, and we plan to present in the future classes of manifolds for which that the singularities of the representation variety play no r\^ole.

\bigskip

Let us now introduce the cohomological language which will allow us to compute the divergence degree of a foam, and eventually relate its dominant part to the Reidemeister torsion. First, let us switch notation and set, for any flat connection $\phi$,
\be \label{notationswitch}
\delta^0_\phi(\Gamma,G) \equiv d\gamma_{\phi|\unit}\qquad \text{and} \qquad\delta_{\phi}^{1}(\Gamma,G) \equiv dH_{\phi}
\ee
for the differentials of $H$ and $\gamma_{\phi}$ at $\phi$ and $\unit\in G$ respectively. Now, a gauge transformation by a group element $h$ changes the holonomies around each face by conjugation,
\be
h\,\cdot\,H_f(A) = h\,H_f(A)\,h^{-1},
\ee
and therefore maps a flat connection to another flat connection. Locally, this means that the directions of the orbit lie in the kernel of $\delta^1_\phi$, and therefore that
\be
\delta^1_\phi(\Gamma,G)\,\circ\,\delta^0_\phi(\Gamma,G) = 0,
\ee
which can be explicitly checked. In other words, the study of flat connections boils down to the cochain complex $C_\phi^*(\Gamma,G)$ defined by
\be
0\longleftarrow C^2_\phi(\Gamma,G) \xleftarrow{\ \delta^1_\phi(\Gamma,G)\ }  C_\phi^1(\Gamma,G) \xleftarrow{\ \delta^0_\phi(\Gamma,G)\ } C_\phi^0(\Gamma,G) \longleftarrow 0,
\ee
where the groups $C^0_\phi$, $C^1_\phi$ and $C^1_\phi$ are respectively $\alg$, $T_\phi G^E$ and $\alg^F$.\footnote{More precisely, the homomorphism $\phi$ turns the Lie algebra $\alg$ into a $\mathbbm{Z}[\pi_1(\Gamma)]$-module denoted $\alg_{\Ad_\phi}$. The complex $C^*_\phi$ is the complex $\alg\otimes_{\pi_1(\Gamma)}C_*(\Gamma,\mathbbm{Z})$, with coefficients in $\alg_{\Ad_\phi}$. Its cohomology $H^*_\phi$ is that of $\Gamma$ with coefficients in $\alg_{\Ad_\phi}$.}
Thus, the tangent to the orbit $\mathcal{O}_\phi$ at $\phi$ is the space of coboundaries $B^1_\phi(\Gamma,G)$, the tangent to  set of non-singular connections $\calF_{0}$ is the space of cocyles $Z^1_\phi(\Gamma,G)$, and the tangent space to the moduli space of non-singular flat connections $\calF_{0}/G$ is $H^1_\phi(\Gamma,G)$. To summarize:
\beq
\begin{aligned}
        T_\phi\calF_0 &= Z^1_\phi(\Gamma,G),\\
        T_\phi\calO_\phi &= B^1_\phi(\Gamma,G),\\
        T_\phi\bigl(\calF_0/G\bigr) &= H^1_\phi(\Gamma,G).
       \end{aligned}
\ee
For notational simplicity, we will often drop hereafter the dependence on $\Gamma$ and $G$ in the twisted cohomology.

\medskip

Note that, because $\calA$ is a Riemannian manifold, all three cochain groups are naturally equipped with inner products. Moreover, it is easy to check that the Euler characteristic $\chi_\phi$ of the twisted cochain complex is actually independent of $\phi$, given by
\be
\chi_\phi = b^2_\phi-b^1_\phi+b^0_\phi = \bigl(\dim G\bigr)\,\chi(\Gamma),
\ee
where $\chi(\Gamma)$ is the Euler characteristic of the cellular homology of $\Gamma$.

\bigskip

We have seen in the section \ref{foams and bubbles} that divergences are to be expected when the faces of $\Gamma$ are not independent (in the sense of the cellular homology), but also more generally when the components of the curvature application $H(\phi)\in\alg^F$ are not independent. Locally, this happens when $H$ is not submersive at $\phi$, i.e. when the rank of $\delta^1_\phi$ is {\it not} $(\dim G)\,F$. We thus introduce the following terminology, standard in differential geometry:
\begin{definition}
A connection $A$ is {\it regular} if $H$ is submersive at $A$. Otherwise it is {\it critical}.
\end{definition}
In particular, regular flat connections on $\Gamma$ are non-singular.
In our cocomplex, the group $H^2_\phi$ counts the 2-cochains which are not in the image of $\delta^1_\phi$: $H^2_\phi = C^2_\phi/B^2_\phi$. A flat connection is thus critical as soon as its second Betti number for the twisted cohomology described above, defined by
\be
b^2_\phi \equiv \dim H^2_\phi = (\dim G)F-\rk\delta^1_\phi,
\ee
is non-zero. Barrett and Naish-Guzman \cite{barrett-PR} have shown that the insertion of a certain tree-like observable in the Ponzano-Regge partition function makes it a well defined distribution on a patch of $\calF_{0}$ where all flat connections are regular. In general, however, we expect the partition function to be controlled by flat non-singular, but \emph{critical} connections, and the divergence degree to be controlled by $b^2_\phi$, according to the intuition that divergences come from linearly dependent components of $dH_\phi$.

Notice that, by definition, the function $\phi\mapsto b^{2}_{\phi}$ is in fact constant on $\calF_{0}$ and equal to its minimum value. We will use the following notation:
\begin{definition}
We denote $b^2_{0}=\min_{\phi\in\calF}b_{\phi}^2$ the constant value of the second twisted Betti number on the space of non-singular flat connections $\calF_{0}$.
\end{definition}
We will show in the next section that (unless singularities spoil the result) this number is indeed the divergence degree of a foam, thereby extending the result of  \cite{barrett-PR} to include critical connections as well as regular ones. Examples where the value of $b^2_0(\Gamma)$ is easily computed will be detailed in the coming sections. Let us mention here the simplest case: when $\phi$ is the trivial connection, one can check immediately that the twisted cohomology reduces to the standard cellular cohomology with coefficients in $\alg$: $C^*_\unit(\Gamma,G)=C^*(\Gamma,\alg)$.\footnote{It this sense, our previous results in \cite{cell-homology} are a particularly simple instance of those presented here.}

\section{Divergence degree away from singularities}\label{divergence}
In this section, we compute the divergence degree of the partition function $\mathcal{Z}(\Gamma,G)$ away from singularities in a regularization of $\mathcal{Z}(\Gamma,G)$ using the heat kernel on $G$. We also study its transformation under Tietze moves, and illustrate our result in the well-known case where $\Gamma$ has the topology of a closed orientable surface.

\subsection{Heat kernel regularization}

The heat kernel on the compact Lie group $G$ is the fundamental solution of the heat equation
\be
\bigl(\partial_\tau - \Delta\bigr)\,\k(g)=0,
\ee
in which $\Delta$ is the Casimir-Laplace operator on $G$. The heat kernel is a central function. Thanks to the Peter-Weyl theorem, it can be decomposed over the characters of $G$, which form an orthonormal basis of eigenfunctions of $\Delta$, with eigenvalues $C(\rho)$:
\be
\k(g) = \sum_{\rho} \bigl(\dim\rho)\,e^{-\tau C(\rho)}\ \chi_\rho(g).
\ee
When $\tau$ goes to zero, this goes to $\sum_\rho(\dim\rho)\chi_\rho(g)=\delta(g)$ which is indeed the usual expansion of $\delta(g)$.

For small times $\tau$, the heat kernel is localized around zero, and its behaviour is close to the Euclidean kernel. In a neighbourhood of the identity,
\be
\k(g) \u{\tau\rightarrow 0}{\sim} \Lambda_{\tau}^{\dim G}\ e^{-\f{|g|^2}{4\tau}},
\ee
where $\lvert g\rvert$ is the Riemannian distance from the identity to $g$, and 
\be
\Lambda_{\tau}\equiv(4\pi\tau)^{-1/2}.
\ee 

We thus define the \emph{regularized} partition function as
\be\label{regularized}
\calZ_\tau(\Gamma,G) \equiv \int_{\cal A}dA\ \prod_{f\in\Gamma_2} \k\bigl(H_f(A)\bigr),
\ee
The integrand is obviously gauge invariant, thanks to the centrality of $\k$. Within this regularization, we can define the degree of divergence in the limit $\tau\rightarrow0$:

\begin{definition}
The {\it divergence degree} of a foam $\Gamma$ is, when it exists, the number $\Omega(\Gamma,G)$ such that the limit $$\mathcal{Z}'(\Gamma,G)\equiv\u{\tau\rightarrow 0}{\lim}\ \Lambda_{\tau}^{-\Omega(\Gamma,G)} \calZ_{\tau}(\Gamma,G)$$ is finite and non-vanishing. In this case, we call this limit the {\it dominant part} of the partition function.
\end{definition}

\subsection{Transformations under changes of presentation of the fundamental group} \label{sec:invariance}

As explained in the beginning of section \ref{sec:twistedhom}, the amplitude $\calZ(\Gamma,G)$ can be seen as a function of a finite group presentation. From this perspective, it is a natural question to ask whether different presentations of the same group yield the same number. It is known that finite presentations of isomorphic groups are related by a finite sequence of two types of elementary transformations, the Tietze moves:

\begin{enumerate}
\item
The Tietze moves of type $1$ consist in the addition of a new generator together with a new relation expressing it as a word in the old generators, and the inverse operation.
\item
The Tietze moves of type $2$ consist in the addition of a new relation which is implied by the old ones, and the inverse operation.

\end{enumerate}

Let us see how the regularized amplitudes $\calZ_\tau(\Gamma,G)$ transforms under them.



On the foam $\Gamma$, the first Tietze move means that we add an edge $e^*$, together with a group element $g_{e^*}$, and a face $f^*$ carrying the relation $g_{e^*} = w(g_1,\dotsc,g_E)$. Quite clearly, since $e^*$ only appears on the boundary of $f^*$, one can use the translation invariance of the Haar measure to show that the regularized amplitude $\calZ_\tau(\Gamma,G)$ is invariant under this operation:
\begin{align}
\int_{G\times G^E} dg_{e^*}\,dA\ K_\tau\bigl(g_{e^*}\,w(g_1,\dotsc,g_E)\mone\bigr)\ \prod_{f\in\Gamma_2}K_\tau(H_f(A)) &= \int_{G} dg\ K_\tau(g)\ \int_{G^E}dA\ \prod_{f\in\Gamma_2}K_\tau(H_f(A)),\\
&= \calZ_\tau(\Gamma,G).
\end{align}
In the last line, we have used the fact that the integral of the heat kernel is normalized to 1 for any $\tau$. The process of adding to $\Gamma$ an edge and a face as described above is called an elementary \emph{2-expansion} of $\Gamma$, while the inverse move is an elementary \emph{2-collapse}. A finite sequence of elementary $2$-expansions and $2$-collapses forms a \emph{2-deformation} of $\Gamma$, and the above argument shows that $\calZ_\tau(\Gamma,G)$, hence (when it exists) the dominant part of the partition function, is invariant under such $2$-deformations, which are particular cases of simple-homotopy equivalences.

The case of Tietze moves of type $2$ is subtler. Consider the first of them, which adds a new, redundant, relation. At the level of the foam, this corresponds to the addition of a new face $f^*$, such that the holonomy around $f^*$, $H_{f^*}(A)$, is automatically trivial when the other flatness conditions hold. An obvious formal manipulation on the \emph{unregularized} amplitude (\ref{amplitude}) suggests that this should change the divergence degree of the foam, \emph{but not the dominant part} of $\calZ(\Gamma,G)$:
\beq \label{tietze 2}
\int_{\calA} dA\ \delta\bigl(H_{f^*}(A)\bigr)\ \prod_{f\in\Gamma_2}\delta(H_f(A))= \delta(\unit)\ \calZ(\Gamma,G).
\ee
For this reason, it has been claimed that $\calZ(\Gamma,G)$ should be a function of $\pi_1(\Gamma)$ and $G$ only \cite{boulatov-model}. But, when applied to the \emph{regularized} amplitude, this transformation yields
\beq \label{tietze 3}
\int_{\calA} dA\ K_\tau\bigl(H_{f^*}(A)\bigr)\ \prod_{f\in\Gamma_2}K_\tau(H_f(A)), 
\ee
which is \emph{not} related to $\calZ_\tau(\Gamma,G)$ in any simple way. The reason is that, although the set of flat connections $\calF\subset \calA$ is unchanged, the rate at which the integrand flows away from $\calF$ is modified. Roughly speaking, the presence of an additional heat kernel makes the integrand more sharply peaked on $\calF$: it decreases the typical width around $\calF$ of the connections contributing to the integral. It follows from this that the dominant part of $\calZ(\Gamma,G)$ is changed by Tietze moves of type $2$, and does depend on the actual presentation of $\pi_1(\Gamma)$. Topologically, the Tietze moves of type 2 generate homotopy equivalences of $2$-complexes which are not \emph{simple}. Like the Reidemeister torsion, to which it is related (sec. \ref{reidemeister}), the dominant part $\calZ'(\Gamma,G)$ may distinguish foams which have different simple-homotopy types, although they are homotopy equivalent.

\subsection{Divergence degree on the space of non-singular flat connections} \label{sec:twisted counting}

Our strategy to deal with the regularized integral (\ref{regularized}) will be the same as the one employed by Forman \cite{Forman-small} to deal with two-dimensional Yang-Mills theory: we will split the integral over $\mathcal{A}$ into an integral over the space of flat connections $\calF$ and another over the normal space to $\calF$. However, like him,\footnote{More exactly, since the exact value of the partition function was known independently thanks to Witten's formula (\ref{sf 2d}), Forman {\it knew} that the singularities did not contribute. In our more general setting, unfortunately, no such result is available.} {\it we will assume that the singularities of $\calF$ do not contribute to the integral}, and thus consider its non-singular, smooth subset $\calF_{0}$ only. Although we do not understand the scope of this assumption in full generality, we will show in the section \ref{YM} how it can be checked in the two-dimensional case. A counter-example in three dimensions is given in the Appendix.

On the manifold of non-singular connections $\calF_{0}$, the following property holds. For each $\phi \in \calF_{0}$, the tangent space splits as
\be\label{split}
T_\phi\mathcal{A} = T_\phi\calF_{0} \oplus N_\phi\calF_{0},
\ee
where $T_\phi\calF_{0}=\ker\delta^1_\phi$ and $N_\phi\calF_{0}=(\ker\,dH_\phi)^\perp$. This property is the relevant condition to apply a generalized Laplace approximation to the integral in (\ref{regularized}). Physically, $T_\phi\calF_{0}$ is a local version of the space of solutions to the flatness equation of motion, while the normal space $N_\phi\calF$ represents the degrees of freedom of a connection which are fixed by the flatness condition, or equivalently which need to be fixed in order to localize the integral on flat connections. As far as divergences are concerned, the intuitive idea is that divergences are likely to occur as soon as the dimension of $N_\phi\calF$ (i.e. the number of degrees of freedom) is smaller than the number of constraints imposing flatness.

Since the heat kernel on $G$ decays exponentially away from the unit, we can begin by restricting the integral (\ref{regularized}) to a tubular neighborhood of $\calF_{0}$, equipped with the normal fibration induced by (\ref{split}). Assuming this neighborhood is small enough, we can then use the exponential mapping to pull the integral in the direction normal to $\calF_{0}$ to the normal spaces $N_{\phi}\calF_{0}$, and consequently write $H(A) = H(\phi, \exp_\phi(y))$, with $y\in N_{\phi}\calF_0$. Denoting $\vol_{\calF_0}$ the volume form on $\calF_{0}$ induced by the Riemannian metric of $G^E$, and $\vol_{N_\phi\calF_{0}}$ the volume form on the normal fibers, the Haar measure on $\calA$ can be written
\be
dA = \vol_{\calF_{0}}\ \vol_{N_\phi\calF_{0}}.
\ee
From this argument, it follows that
\begin{align}
\calZ_\tau(\Gamma,G)& = \int_{\calA} dA\ \prod_{f\in\Gamma_2} \k\bigl(H_f(A)\bigr), \\
&\u{\tau\rightarrow 0}{\sim}\Lambda_{\tau}^{(\dim G)F}\int_{\calF_{0}}\vol_{\calF_{0}}\int_{N_\phi\calF_{0}} \vol_{N_\phi\calF_{0}}\ \exp\Biggl(-\sum_{f\in\Gamma_2}\f{\lvert H_f(\phi,\exp_\phi(y))\rvert^2}{4\tau}\Biggr).
\end{align}
In order to take advantage of the Gaussian behaviour of the heat kernel $\k$ at small times $\tau$, the Riemannian distance between $H_f(\phi,\exp_\phi(y))$ and the identity can be expanded around the flat configuration $\phi$. This gives
\be
\lvert H_f(\phi,\exp_\phi(y))\rvert^2 =  \Vert dH_{f|\phi}(y)\Vert_{\alg}^2 + \mathcal{O}(y^3),
\ee
It should be emphasized that the differential of $H_f$ is only contracted with the variables $y$ of the normal space at $\phi$.

We can then perform the Gaussian integral over the fibers, choosing an arbitrary basis $d^1_\phi=(d^1_{\phi,\alpha})_\alpha$ of $N_\phi\calF_{0}$, and write $y=y^\alpha d^1_{\phi,\alpha}$ so as to integrate over the variables $y^\alpha$. This results in the following expression, whose terms shall be explained hereafter,
\be \label{gaussian int}
\int_{N_\phi\calF_{0}} \vol_{N_\phi\calF_{0}}\ \exp\Biggl(-\f{\Vert dH_{\phi}(y)\Vert_{\alg^{F}}^2}{4\tau}\Biggr) = \Lambda_{\tau}^{-\dim(\ker dH_\phi)^\perp}\ \f{\vol_{N_\phi\calF_{0}}(d^1_\phi)}{\vol_{B^2_\phi}(dH_\phi(d^1_\phi))}.
\ee
First, $\vol_{N_\phi\calF_{0}}(d^1_\phi)$ denotes the volume spanned by the vectors of the basis $d^1_\phi$ in the tangent space $N_\phi\calF_{0}$. It comes from evaluating the volume form $\vol_{N_\phi\calF_{0}}$ at the saddle point $y=0$ in the basis $d^1_\phi$. Second, as the Gaussian integral is performed with respect to the variables $y^\alpha$, the square root of the determinant of the Hessian $\Vert dH_{\phi}(y)\Vert_{\alg^{F}}^2$ in the basis $d^1_\phi$ appears. In geometric terms, this is the volume spanned by the vectors $(dH_\phi(d^1_\phi{\alpha}))_\alpha$ in the image $B^2_\phi$. Third, notice that the right hand side of \eqref{gaussian int} is naturally independent of the choice of $d^1_\phi$.  We give some further details on how to compute these quantities in practice in the next section \ref{YM}.

Coming back to the full expression of $Z_\tau(\Gamma, G)$, the dependence on $\tau$ is now extracted from the integrals, and we are left with
\be \label{amplitude scaling}
\calZ_\tau(\Gamma,G) \u{\tau\rightarrow 0}{\sim}\Lambda_{\tau}^{\Omega(\Gamma,G)} \int_{\calF_{0}}
\f{\vol_{N_\phi\calF_{0}}(d^1_\phi)}{\vol_{B^2_\phi}\bigl(dH_\phi(d^1_\phi)\bigr)}\ \vol_{\calF_{0}}.
\ee
Collecting the exponents of $\tau$ coming both from the asymptotics of the heat kernel and from the Gaussian integral, the degree of divergence is $\Omega(\Gamma,G) = (\dim G)F-\dim(\ker dH_{\phi})^\perp$, with the right hand side computed on non-singular flat connections. To make the link with the twisted cohomology previously introduced, recall that $dH_{\phi}$ is the coboundary operator $\delta^1_{\phi}$, hence the dimension of $(\ker dH_{\phi})^\perp$ is the rank of $\delta^1_{\phi}$. Since $\bigl((\dim G)F - \rk\delta^1_{\phi}\bigr)=b^2_{0}$, we have proved that
\begin{theorem}
Whenever the ratio in (\ref{amplitude scaling}) is integrable with respect to the Riemannian volume form on the set of non-singular flat connections $\calF_{0}$, the divergence degree of a closed foam is given by
$$
\Omega(\Gamma,G) = b^2_{0}(\Gamma,G),
$$
where $b^2_{0}(\Gamma,G)$ is the value of the second twisted Betti number on $\calF_{0}$.
\end{theorem}
This is the main result of this paper, and can be read as a precise realization of the idea that divergences may occur when the rank of the linearized system of flatness constraints is smaller than the number of constraints. Let us now illustrate this result in the two-dimensional case.

\subsection{$\SU(2)$ Yang-Mills theory on closed orientable surfaces} \label{YM}

The case when $\Gamma$ is the cell decomposition of an orientable closed surface is well-known: it is the weak-coupling (or small volume) limit of two dimensional Yang-Mills theory \cite{Forman-small}. Since the seminal works of Atiyah-Bott \cite{atiyah-bott-2d-YM}, Goldman \cite{goldman} and Witten \cite{witten-2d-YM}, the structure of the moduli space of flat connections has been comprehensively studied, and elucidated. Here, we illustrate how to use the local twisted cohomology to extract divergence degrees. For concreteness, we set $G=\SU(2)$ in this section. We show that there are no divergence for surfaces of genus greater than 2. In addition we treat the special case of the 2-torus, where non-singular flat connections are critical: we compute its divergence degree away from singularities, and explain why the latter do not affect the powercounting result.

\begin{theorem}
If the closed foam $\Gamma_g$ is the standard cell decomposition of a closed orientable surface $\Sigma_g$ of genus $g$ with $2g$ edges and one face (see below), its global degree of divergence is given by 
$$
\Omega(\Gamma_g,\SU(2))= b^2_0(\Gamma_g,\SU(2)) \,=\, \begin{cases}
3& \text{if $g=0$} \\
1& \text{if $g=1$} \\
0 & \text{if $g\geq 2$}.
\end{cases}
$$
\end{theorem}

The simply connected case, $g=0$, is somewhat easier. We have shown indeed in \cite{cell-homology} that for any simply connected cell 2-complex $\Gamma$, the degree of divergence is given by $\Omega(\Gamma,G) = (\dim G)\,b^2(\Gamma)$, where $b^2(\Gamma)$ is the second Betti number in the standard cellular cohomology of $\Gamma$. Moreover, it is already known for $g\geq 2$ that $\lim_{\tau\rightarrow 0}\mathcal{Z}_{\tau}(\Gamma_{g},\SU(2))$ is finite, and is given by the integral of the combinatorial Reidemeister torsion over the moduli space $\calF_{0}/\SU(2)$ \cite{witten-2d-YM}. (We will show in the next section that this is still true for a generic foam -- again, under the assumption that the singularities of the torsion are integrable.)

 Let us consider the cases $g\geq 1$ in more detail. As well-known, a cell decomposition of an orientable close 2d surface of genus $g$ can be reduced to a flower graph with $2g$ edges supporting only one face. This corresponds to the following presentation of the fundamental group:
\be
\pi_1(\Gamma_g) = \langle a_1,\,b_1,\dotsc,a_g,\,b_g\ \vert\ [a_1,b_1]\dotsm [a_g,b_g]=\unit \rangle.
\ee
It has $2g$ generators, and one relation. Here the square brackets denote the group commutator in $\pi_1(\Gamma_g)$: $[a,b] \equiv aba^{-1}b^{-1}$. The space of discrete $\SU(2)$-connections is simply obtained by mapping the generators to elements of $\SU(2)$, while the relator corresponds to the flatness condition:
\be
{\calF}_g \equiv \bigl\{ (a_1,\,b_{1},\dotsc,a_g,\,b_g)\ \in\,\SU(2)^{2g},\ [a_1,b_1]\dotsm [a_g,b_g]=\unit \bigr\}.
\ee

In this two-dimensional case, it is not difficult to convince oneself that there are singular connections in ${\calF}_g$, as the differential of the flatness relation is clearly not of constant rank. This means that the assumptions used in section \ref{sec:twisted counting} fail to be satisfied globally on $\calF_g$. Also, different isotropy groups for the action of gauge transformations are involved, making the moduli space $\calF_g/\SU(2)$ ill-defined as a manifold. But of course, the reasoning of the section \ref{sec:twisted counting} still applies on the smooth set of non-singular representations of $\pi_1(\Gamma_g)$ into $\SU(2)$.


Actually, this two-dimensional situation admits a specific structure which simplifies the analysis: the non-singular flat connections are exactly the irreducible ones. We know that a $G$-manifold is stratified by the action of $G$ (and admits an open and dense principal stratum). Here, it means that $\SU(2)^{2g}$ decomposes into a finite number of smooth manifolds, the strata, each of them consisting in the set of points whose isotropy group is conjugated to a given subgroup $H$ of $\SU(2)$. Then, the simplification comes from Poincar\'e duality which gives in 2d:
\be
H^0_\phi \simeq H^2_\phi,
\ee
for any flat connection $\phi$ \cite{goldman}. This means that we have a stratification of $\calF_g$ according to the different possible isotropy groups, coinciding with a stratification of $\calF_g$ according to the rank of the coboundary operator $\delta^1_\phi$ (since $\rk \delta^1_\phi = (\dim G)-b^2_\phi$). In particular, the irreducible flat connections, when they exist, are regular (and thus non-singular), whereas reducible flat connections are critical.

To be more explicit, let us distinguish the cases $g\geq 2$ and $g=1$, the latter being somehow more singular.
\begin{itemize}
\item{$g\geq 2$} - We can write:
\be
\calF_{g\geq 2} = \calF_0 \cup \calF_{(T)} \cup \calF_{(G)},
\ee
where $\calF_0$ is the principal stratum:
\be
\calF_0 = \bigl\{ \phi\in\calF,\, \zeta(\phi)=\zeta(G)\bigr\} = \bigl\{ \phi\in\calF,\, \rk \delta_\phi^1 = \dim G = 3\bigr\}.
\ee
Its stabilizer is the center $\zeta(G)$ of $G$ and the orbits are isomorphic to $G/\zeta(G)$. The space $\calF_{(H)}$ for $H=T,G$ are smooth manifolds for which the isotropy group $H$ is either the torus $T = \U(1)$ or $G$ itself:
\be
\calF_{(H)} = \bigl\{ \phi\in\calF,\, \zeta(\phi)\simeq H \bigr\} = \bigl\{ \phi\in\calF,\, \rk\delta^1_\phi = \dim G/H \bigr\}.
\ee
Now let us see how this can be used in more details. Let us first show that there are indeed non-singular connections where the rank of $\delta^1_\phi$ is 3, so that $\calF_0$ is not empty. To this purpose, pick elements $a_i,b_i\in\SU(2)$ trivializing the commutators $[a_i,b_i]=\unit$ for $i=1,\dotsc,g$. The group elements $a_i$ and $b_i$ should thus have the same axis of rotation $\hat{n}_i\in S^2$, but can have different axes for different $i$. Pulling $\delta^1_\phi$ back to the unit so that it acts on the Lie algebra $\alg^E = T_{{\unit}}G^E$, we have:
\be
\delta^1_\phi(u,v) = \sum_{i=1}^g \bigl(1-\Ad_{b_i}\bigr)\,u_i - \bigl(1-\Ad_{a_i}\bigr)\,v_i.
\ee
for algebra elements $u\equiv(u_i)_{i=1,\dotsc,g}$ and $v\equiv(v_i)_{i=1,\dotsc,g}$. Identifying $u_i$ and $v_i$ with 3-vectors, the adjoint action $\Ad_{a_i}$ is just a rotation around the axis $\hat{n}_i$. Thus for each $i$, it is natural to decompose $u_i$ and $v_i$ into orthogonal and parallel components to the direction $\hat{n}_i$: $u_i = u_{i\parallel}+u_{i\perp}$ and similarly for $v_i$. Then all parallel components disappear, since $(1-\Ad_{a_i})v_i = (1-\Ad_{a_i})v_{i\perp}$. Moreover, the latter quantity still belongs to the orthogonal plane to $\hat{n}_i$, where $(1-\Ad)$ is invertible. This means that, varying $u_i$ and $v_i$, each term of the sum in the above equation spans the orthogonal plane to $\hat{n}_i$. Finally, if at least two axes among the $g$ directions are distinct, then the span of $\delta^1_\phi(u_i,v_i)$ is the whole algebra $\alg$. So there exist flat connections where the curvature map is submersive, i.e. the set of (non-singular) regular flat connections is not empty. Consequently, by the implicit function theorem, the latter form a smooth manifold of dimension
\be
\dim \calF_0 = \dim \ker\delta^1_{\calF_0} = 6g - 3,
\ee
and $b^2_{0}(\Gamma_g,\SU(2)) = 0$. Then, the reasoning of section \ref{sec:twisted counting} leads to the conclusion of a finite result away from singularities. We refer to the work of Sengupta \cite{sengupta-singularities-2d} for details on why singularities do not contribute to the final result. Let us simply say that the Abelian connections correspond to taking the same axis $\hat{n}$ for all group elements, so that $\calF_{(T)}$ is a manifold of dimension $2g+2$, on which $\rk\delta^1_\phi = 2$. In particular, observe that the relation $\dim \calF = \dim \ker \delta^1_\phi$ does not apply to the singular strata.

The most efficient way to compute the partition function in this situation is to expand the heat kernel onto representations and integrate the group elements using the orthogonality relation of matrix elements:
\be
Z_\tau(\Gamma_g,\SU(2)) = \int \prod_{i=1}^g da_i\,db_i\ \k\bigl([a_1,b_{1}]\dotsm [a_g,b_g]\bigr)= \bigl(\Vol G\bigr)^{2g} \sum_{j\in\f{\mathbbm{N}}{2}} \f{e^{-\tau j(j+1)}}{(2j+1)^{2g-2}}.
\ee
It is well-defined for $g\geq 2$ when $\tau$ goes to zero:
\be
Z_{\tau=0}(\Gamma_g,\SU(2)) = \bigl(\Vol G\bigr)^{2g} \sum_{n\geq 1} n^{-(2g-2)}.
\ee

\item{The torus case, $g=1$} - Now we look in details at the case of the 2-torus, which is manifestly divergent if we try to use the above formula directly. The reason is that the isotropy group of any flat connection never reduces to the center of $G$, and is at least a $\U(1)$ torus. Through Poincar\'e duality, this implies that the curvature map is not submersive on flat connections, so that non-singular flat connections are critical, i.e. $H^2_\phi(\Gamma_{g=1})\neq 0$. Now, let us show that the the result of the section \ref{sec:twisted counting} apply nevertheless. The curvature map is here a single group commutator, $(a,b)\mapsto [a,b]=aba\mone b\mone$, so that flat connections consist in rotations with the same axis $\hat{n}\in S^2$:
\be \label{flat connections torus}
\calF_{g=1} = \bigl\{ (a,b) \in \SU(2)^2,\ a = \exp(i\psi_a\, \hat{n}\cdot\vec{\sigma}),\ b = \exp(\pm i\psi_b\, \hat{n}\cdot\vec{\sigma}),\ \text{with}\ \hat{n}\in S^2, (\psi_a,\psi_b)\in[0,\pi)^2 \bigr\}.
\ee
Notice the `$\pm$' in $b$ due to the Weyl symmetry. Clearly, $\calF$ is a smooth manifold of dimension 4. The key point is that the tangent space to $\calF$ is given by the kernel of $\delta^1_\phi$ whenever either $a$ or $b$ (say $b$) is not in the center of $G$. Indeed,
\be
\delta^1_\phi = d\bigl([a,b]\bigr) = \bigl(1-\Ad_b\bigr)\,da\,a^{-1} - \bigl(1-\Ad_a\bigr)\,db\,b^{-1}.
\ee
The operators $(1-\Ad_a)$ and $(1-\Ad_b)$, seen as linear maps on $\alg\simeq\mathbb{R}^3$, have the same one-dimensional kernel, the direction parallel to $\hat{n}$: this corresponds to variations of the angles $\psi_a$, $\psi_b$ for a fixed $\hat{n}$. Restricted to the orthogonal plane, $(1-\Ad_b)$ is invertible, since $\Ad_b$ is a non-trivial rotation. Thus, the equation for the kernel of $\delta^1_\phi$ reads 
\beq
u_{a\perp} = (1-\Ad_b)^{-1} (1-\Ad_a) u_{b\perp},
\ee
and therefore fixes $2$ real components of $(u_a,u_b)\in\mathbb{R}^3\times\mathbb{R}^3$ (i.e. $\rk\delta_\phi^1=2$). It expresses the condition that the connection remains flat under variations of the directions of $a$ and $b$. We can thus evaluate the second twisted Betti number on the set of non-singular flat connections: since the coboundary operator $\delta^1_\phi$ is of rank 2, we have $b^2_0 = 3-2=1$. 

But again, let us stress the existence of singularities, although $\calF$ is smooth: if $a,b\in\zeta(G)$, $\delta^1_\phi$ is the zero map. There it is clear that the tangent space to $\calF$ is not the kernel of $\delta^1_\phi$: $T_\phi\calF\simeq \mathbb{R}^4 \neq \ker\delta_\phi^1=T_\phi \SU(2)^2$. These situations correspond to an isotropy group which is $G$ itself, and we have the stratification
\be
\calF_{g=1} = \calF_{(\textrm{U}(1))} \cup \calF_{(G)},
\ee
with $\calF_{(G)}=\zeta(G)^2\subset \overline{\calF_{(\textrm{U}(1))}}$.

Let us now apply the method described in the previous section to compute the small $\tau$ behaviour of the partition function
\be
Z_\tau(\Gamma_{g=1},\SU(2)) = \int_{\SU(2)^2} da\,db\ \k\bigl([a,b]\bigr).
\ee
We parametrize the directions $\hat{n}_a$ and $\hat{n}_b$ with spherical angles $(\theta_a,\varphi_a)$ and $(\theta_b,\varphi_b)$, so that the Haar measure is $da = \sin^2\psi_a\,\sin\theta_a\,d\psi_a\,d\theta_a\,d\varphi_a$, and similarly for $db$. The saddle points $\phi\in\calF$ are simply given by $\hat{n}_b = \pm\hat{n}_a$, and for now we focus on the $\hat{n}_b = +\hat{n}_a$ component of $\calF$. To use our formula \eqref{amplitude scaling}, we need to find a basis $d^1_\phi$ of the orthocomplement of $\ker\delta^1_\phi$. First, it is convenient to change basis in the Lie algebra from the standard Cartesian basis $i\vec{\sigma}=(i\sigma_x,i\sigma_y,i\sigma_z)$ to the spherical basis: $\tau_n\equiv \hat{n}_a\cdot i\vec{\sigma}$, $\tau_\theta \equiv \cos\theta_a(\cos\varphi_ai\sigma_x+\sin\varphi_ai\sigma_y)-\sin\theta_ai\sigma_z$, and $\tau_\varphi\equiv -\sin\varphi_ai\sigma_x+\cos\varphi_ai\sigma_y$. Then, the Maurer-Cartan 1-form reads
\be
da\,a^{-1} = \tau_n\,d\psi_a + \sin\psi_a\, \bigl(\cos\psi_a\, \tau_{\theta} - \sin\psi_a\, \tau_\varphi\bigr)\,d\theta_a + \sin\psi_a\,\bigl(\sin\psi_a\, \tau_\theta + \cos\psi_a\, \tau_\varphi\bigr)\,\sin\theta_a\,d\varphi_a,
\ee
and similarly for $db\,b\mone$. Using this expression, we can compute the action of $\delta^1_\phi$ on tangent vectors
\be
y\equiv y^{\theta_a}\pp_{\theta_a}+\f{y^{\varphi_a}}{\sin\theta}\pp_{\varphi_a}+y^{\theta_b}\pp_{\theta_b}+\f{y^{\varphi_b}}{\sin\theta}\pp_{\varphi_b},
\ee
with the vector fields evaluated on $\calF$. Here, $\theta$ is the common value of $\theta_a$ and $\theta_b$ on $\calF$. This gives 
\begin{multline}\label{delta1torus}
\delta^1_\phi(y) = 2\sin\psi_a\sin\psi_b\Biggl\{\Bigl(\sin(\psi_a+\psi_b)\,\tau_\theta + \cos(\psi_a+\psi_b)\,\tau_\varphi\Bigr)\,\bigl(y^{\theta_a}-y^{\theta_b}\bigr)\\
+ \Bigl(\sin(\psi_a+\psi_b)\,\tau_\varphi - \cos(\psi_a+\psi_b)\,\tau_\theta\Bigr)\,\bigl(y^{\varphi_a}-y^{\varphi_b}\bigr)\Biggr\}.
\end{multline}
This is a rather simple operator, namely the composition of a rotation and a homothety. Its kernel is generated by the vectors $(\partial_{\theta_a}+\partial_{\theta_b})$ and $(\partial_{\varphi_a}+\partial_{\varphi_b})$, in accordance with the fact that the variations of the directions $\hat{n}_a$, $\hat{n}_b$ must be identical in order for the connection to stay on $\calF$. An orthonormal basis of $(\ker\delta^1_\phi)^\perp$ is then
\beq
d^1_\phi \equiv \left( \f{\sin\psi_a\,\sin\psi_b}{\sqrt{\sin^2\psi_a + \sin^2\psi_b}}\left( \f{1}{\sin^2\psi_a}\partial_{\theta_a} - \f{1}{\sin^2\psi_b}\partial_{\theta_b}\right),\ \f{\sin\psi_a\,\sin\psi_b}{\sin\theta\sqrt{\sin^2\psi_a + \sin^2\psi_b}}\left(\f{1}{\sin^2\psi_a}\partial_{\varphi_a} - \f{1}{\sin^2\psi_b}\partial_{\varphi_b} \right)\right).
\ee
From this, we can compute the volumes entering our formula \eqref{amplitude scaling} for non-singular flat connections:
\begin{align}
&\vol_{N_\phi\calF}(d^1_\phi) = 1, \\
&\vol_{B^2_\phi}(\delta^1_\phi(d^1_\phi)) = 4\,\bigl(\sin^2\psi_a + \sin^2\psi_b\bigr).\label{gaussian int 2d}
\end{align}
As we anticipated, the determinant coming from the Gaussian integral \eqref{gaussian int 2d} is singular when $a$ and $b$ both approach the center of $G$, where $\sin\psi_a=\sin\psi_b=0$. However, this singularity is cancelled by the induced Riemannian volume form on $\calF$
\beq
\vol_{\calF} = \bigl(\sin^2\psi_a + \sin^2\psi_b\bigr)\ d\psi_a\,d\psi_b\ \sin\theta\,d\theta\,d\varphi,
\ee
Moreover, gauge transformations leave the class angles $\psi_a, \psi_b$ invariant, and act as rotations of the direction $\hat{n}=(\theta, \varphi)$. It follows that
\be
Z_\tau(\Gamma_{g=1},\SU(2)) \u{\tau\rightarrow 0}{\sim} \f{2}{4}\,\Lambda_\tau \int_{S^2}\vol_{S^2}\ \int_{[0,\pi]^2} d\psi_a\,d\psi_b= 2\pi\,\Lambda_{\tau}\ \int_{[0,\pi]^2} d\psi_a\,d\psi_b.
\ee
Note that we have multiplied the whole expression by a factor $2$ to take into account the set of flat connections where $\hat{n}_b=-\hat{n}_a$ instead of $\hat{n}_b=\hat{n}_a$, which we assumed implicitely above. 

Following Witten \cite{witten-2d-YM}, this shows that the symplectic form on the moduli space of flat $\SU(2)$ connections on the 2-torus is just $d\psi_a\wedge d\psi_b$. Moreover, the comparison with the prediction coming from the twisted cohomology $H^*_{\phi}$ is successful, since we have found $\Omega(\Gamma_{g=1},\SU(2))=1=b^2_0(\Gamma_{g=1},\SU(2))$. This happens in spite of the vanishing of the determinant of the Gaussian form in \eqref{gaussian int 2d}, which is compensated by the measure $\vol_{N_\phi\calF}(d^{1}_\phi)\,\vol_\calF$.
\end{itemize}

\section{Relation to Reidemeister torsion} \label{reidemeister}

Our main result, \eqref{amplitude scaling}, shows a factorisation of a regulator-dependent part from an integral which is independent of $\tau$. Thus, our powercounting is true as soon as the integral gives a finite number. This may not be the case due to the singularities of $\calF$, i.e. points where $\dim\ker\delta^1_\phi>\dim\calF$. Thus, we would like to look a bit more precisely at this integral, and try to relate it some known quantities. We know from Witten \cite{witten-2d-YM} that on a closed orientable surface, it is the integral of the Reidemeister torsion for the twisted cohomology. In three dimensions, Witten has shown that the partition function for Riemannian quantum gravity (with degenerate metrics), of which the Ponzano-Regge model \eqref{pr model} is the spinfoam quantization, can be cast into the integral of the Ray-Singer torsion of the spacetime 3-manifold \cite{witten-3d}. As the analytic Ray-Singer torsion is the same as the combinatorial Reidemeister torsion, we expect at the discrete level that the integral over flat connections can be reduced to the integral of the Reidemeister torsion. This is what has been achieved by Barrett and Naish-Guzman in \cite{barrett-PR} when the Ponzano-Regge amplitude is finite. Let us also mention the work of Gegenberg and Kunstatter \cite{gegenberg-partition-function-bf}, which shows the relation between the partition function of the topological BF field theory in any dimension with the torsion of the spacetime manifold. Here, we reach a similar conclusion, but as we do not have spacetime manifolds, we find that the relevant invariant is the Reidemeister torsion of the cell 2-complex $\Gamma$, which is indeed an invariant of simple-homotopy of $\Gamma$. To prove this result, we will assume that {\it all non-singular connections have the same isotropy type}.

As for the problem of the singularities of $\calF$ and the possibility that the integral be infinite, this is related to the difficulty of integrating the torsion in the generic case. This problem is known in mathematics, and at the present day only limited results have been obtained \cite{dubois private}.

\subsection{Extracting the torsion}

We first introduce a convenient and standard notation. If $b$ and $b'$ are two bases of a given vector space related by the matrix $M$: $b'_\alpha = M_\alpha^{\phantom{\alpha}\beta}b_\beta$, we denote the determinant of $M$ by $[b'/b]$.

Pick up a basis $c^2_\phi$ of $C^2_\phi=\alg^F$. Then we identify the space $H^2_\phi$ with the orthocomplement of the image of $\delta^1_\phi$ in $\alg^F$:
\be
H^2_\phi = \alg^F/B^2_\phi \simeq \bigl(B^2_\phi\bigr)^\perp,
\ee
and choose a basis $h^2_\phi$ of $(B^2_\phi)^\perp$. From $h^2_\phi$ and the image of the basis $d^1_\phi$ via $\delta^1_\phi$, we obtain another basis of $C^2_\phi$ written $\delta^1_\phi(d^1_\phi)\,h^2_\phi$. The change of basis is given by
\be
\tau^2_\phi \equiv [\delta^1_\phi(d^1_\phi)\,h^2_\phi/c^2_\phi],
\ee
Thus, to compute the volume spanned by $\delta^1_\phi$, we can first compute the volume spanned by $c^2_\phi$ in $\alg^F$, change basis to $\delta^1_\phi(d^1_\phi)h^2_\phi$ and then divide by the volume of $h^2_\phi$:
\be
\f{1}{\vol_{B^2_\phi}\bigl(\delta^1_\phi(d^1_\phi)\bigr)} = \f{1}{\tau^2_\phi}\ \f{\vol_{(B^2_\phi)^\perp}(h^2_\phi)}{\vol_{\alg^F}(c^2_\phi)}.
\ee
If $\phi$ is regular, i.e. if $H^2_\phi=0$, then there is no $h^2_\phi$, and $\vol_{(B^2_\phi)^\perp}(h^2_\phi)$ should be replaced with 1 in the above formula (see \cite{barrett-PR}).

Let us now describe more carefully the space of flat connections, with the motive of integrating the orbits of the group action. We assume that the isotropy groups of all non-singular flat connections are isomorphic, turning $\calF_{0}$ into a fiber bundle. The isotropy group $\zeta(\phi)$ of $\phi$ is generated by $\ker\delta^0_\phi$. The orbit $\mathcal{O}_\phi$ through $\phi$ is isomorphic to $G/\zeta(\phi)$ and its tangent space is the image of $\delta^0_\phi$: $T_\phi\mathcal{O}_\phi=B^1_\phi$. We now rewrite the volume form on $\calF_{0}$ in order to make the splitting between the orbit directions and their orthogonal directions in $T_\phi\calF_{0}$ explicit. We pick on the one hand a basis $c^1_\phi$ of $C^1_\phi=T_\phi G^E$, and on the other hand we complete the basis $d^1_\phi$ to get a second basis on $C^1_\phi$. This can done by considering a basis $d^{0}_\phi$ of $(Z^{0}_{\phi})^{\perp}$, and its pushforward $\delta_{\phi}^0(d^{0}_\phi)$ in $B^1_\phi$. Let then $h^1_\phi$ be a basis of the first cohomology space $H^1_\phi = T_\phi\calF_{0}/T_\phi\mathcal{O}_\phi$. By identifying $H^1_\phi$ with the orthocomplement of $B^1_\phi$ in $T_\phi\calF_{0}$, we can lift $h^1_\phi$ to a basis of $(B^1_\phi\oplus Z_\phi^1)^\perp$. This way we get a basis, $\delta_{\phi}^{0}(d^0_\phi)\,h^1_\phi\,d^1_\phi$, of $C^1_\phi$ corresponding to the decomposition
\be
T_\phi\,G^E = B^1_\phi\oplus (B^1_\phi\oplus Z^1_\phi)^\perp\oplus N_\phi\calF_{0},
\ee
which can be compared to $c^1_\phi$ through the determinant
\be
\tau^{1}_\phi\equiv[\delta_{\phi}^{0}(d^0_\phi)\,h^1_\phi\,d^1_\phi/c^1_\phi].
\ee
Furthermore, let $\delta_{\phi}^{0}(d^{0}_\phi)^*$ and $(h^1_\phi)^*$ be the dual bases to $\delta_{\phi}^{0}(d^{0}_\phi)$ and $h^1_\phi$. They induce volume forms on the orbits and on the moduli space, denoted $\bigwedge\delta_{\phi}^{0}(d^{0}_\phi)^*$ and $\bigwedge(h^1_\phi)^*$ respectively. Equipped with these bases, we can rewrite the volume form on $\calF_{0}$ as
\be
\vol_{\calF_{0}} = \tau^{1}_\phi\ \f{\vol_{T_\phi G^E}(c^1_\phi)}{\vol_{N_\phi\calF_{0}}(d^1_\phi)}\bigwedge\delta_{\phi}^{0}(d^{0}_\phi)^*\,\bigwedge(h^1_\phi)^*.
\ee
Here, as before, the quantity $\vol_{T_\phi G^E}(c^1_\phi)$ refers to the volume spanned by $c^1_\phi$ in the tangent space $T_\phi G^E$. This results in the following expression for the dominant part of the partition function:
\be \label{finite part F0}
\calZ'(\Gamma,G) = \int_{\calF_{0}} \f{\tau^{1}_\phi}{\tau^{2}_\phi}\ \f{\vol_{T_\phi G^E}(c^1_\phi)\,\vol_{(B^2_\phi)^\perp}(h^2_\phi)}{\vol_{\alg^F}(c^2_\phi)}\ \bigwedge\delta_{\phi}^{0}(d^{0}_\phi)^*\,\bigwedge(h^1_\phi)^*.
\ee

To simplify the following discussion, we will distinguish the reducible and irreducible cases, although the latter can be read as a particular case of the former. 

\subsection{Irreducible connections}

In this case, the stabilizer is the center of the group, $\zeta(\phi) \simeq \zeta(G)$, and via the gauge transformation map $\gamma_{\phi}$ defined in \eqref{gauge map} the orbits $\mathcal{O}_\phi$ are isomorphic to $G/\zeta(G)$. Hence, we can pull back the integral over each orbit to an integral over $G/\zeta(G)$. Since the integrand in \eqref{finite part F0} is gauge invariant, it is sufficient to evaluate the Jacobian of the change of variables at the unit in $G/\zeta(G)$. There, the kernel $Z^0_\phi$ of the linearized gauge transformation operator $\delta_\phi^0$ is zero, and $d^0_\phi$ forms a basis of the full Lie algebra $C^0_\phi=\alg$. Moreover, the volume form on the orbits $\bigwedge\delta_{\phi}^{0}(d^{0}_\phi)^*$ is the push forward of the volume form $\bigwedge(d^0_\phi)^*$ induced on $\alg$, which itself can be related to the Riemannian volume form $\vol_{\alg/\zeta(G)}$ according to
\be
\bigwedge(d^0_\phi)^*=\f{\vol_{G/\zeta(G)}}{\vol_{\alg}(d^0_\phi)}=\f{1}{\tau^0_\phi}\ \f{\vol_{G/\zeta(G)}}{\vol_\alg(c^0_\phi)},
\ee
where in the last equality we introduced a basis $c^0_\phi$ of $\alg$ and the corresponding determinant $\tau^0_\phi\equiv[d^0_\phi/c^0_\phi]$. Hence, 
\be
\calZ'(\Gamma,G) =\int_{\calF_{0}/G}\left(\int_{G/\zeta(G)}\vol_{G/\zeta(G)}\right)\f{1}{\tau_\phi^0\,\vol_\alg(c^0_\phi)}\ \f{\tau^1_\phi}{\tau^2_\phi}\ \f{\vol_{T_\phi G^E}(c^1_\phi)\,\vol_{(B^2_\phi)^\perp}(h^2_\phi)}{\vol_{\alg^F}(c^2_\phi)}\,\bigwedge(h^1_\phi)^*.
\ee
At this point we can identify the Reidemeister torsion of the cochain complex $C^*_{\phi}$
\be
\tor_\phi \equiv\f{\tau^1_\phi}{\tau^0_\phi\tau^2_\phi},
\ee
which by construction depends on the the bases $c^k_\phi$ of $C^k_\phi$ and $h^k_\phi$ of $H^k_\phi$, but not on the bases $d^k_\phi$. As well-known, it defines an invariant of simple-homotopy equivalences of the $2$-complex $\Gamma$ \cite{turaev-book}.

Finally, the integral over the orbit $G/\zeta(G)$ simply gives a factor $\Vol(G)/\#\zeta(G)$, and we obtain
\be
\calZ'(\Gamma,G) =\f{\Vol(G)}{\#\zeta(G)} \int_{\calF_0/G} \tor_\phi\ \f{\vol_{T_\phi G^E}(c^1_\phi)\,\vol_{(B^2_\phi)^\perp}(h^2_\phi)}{\vol_\alg(c^0_\phi)\,\vol_{\alg^F}(c^2_\phi)}\ \bigwedge(h^1_\phi)^*.
\ee
The integrand is naturally independent of any choice of basis. The last step to simplify this expression is borrowed to \cite{barrett-PR}, and consists in choosing the bases $c^k_\phi$ in a suitable way. Let $l$ be a basis of the Lie algebra $\alg$, and  choose $c^0_\phi=l$ together with $c^2_\phi$ made by one copy of $l$ for each face. This implies $\vol_{\alg^F}(c^2) = (\vol_\alg(l))^F$. From $l$, one also gets a basis of $\alg^E$ with one copy of $l$ for each edge. Then push forward this basis isometrically to $C^1_{\phi}=T_\phi G^E$, and choose $c^1_\phi$ as the result of this operation, so that $\vol_{T_\phi G^E}(c^1_\phi) = (\vol_{\alg}(l))^E$. Since the Euler characteristic of the cell-complex $\Gamma$ is $\chi(\Gamma) = F-E+1$, this gives
\be
\calZ'(\Gamma,G) = \f{\Vol(G)}{\#\zeta(G)}\ \bigl(\vol_\alg(l)\bigr)^{-\chi(\Gamma)} \int_{\calF_0/G} \tor_\phi\ \vol_{(B^2_\phi)^\perp}(h^2_\phi)\ \bigwedge(h^1_\phi)^*.
\ee
A particularly simple choice for $l$, of course, would be to pick an orthonormal basis, in which case $\vol_{\alg}(l)=1$. Moreover, we are free to choose for $h^2_\phi$ orthonormal as well, so that $\vol_{(B^2_\phi)^\perp}(h^2_\phi)=1$, and thus
\be\label{finalirreducible}
\calZ'(\Gamma,G) = \f{\Vol(G)}{\#\zeta(G)}\ \int_{\calF_0/G} \tor_\phi\ \bigwedge(h^1_\phi)^*.
\ee
This concludes our argument in the irreducible case.

\subsection{Reducible connections}

The reducible case proceeds along the very same lines, except for one complication: the presence of non-trivial cohomology classes in $H_\phi^0$. To obtain a basis of the Lie algebra $C^0_\phi$, we should therefore pick a basis $h^0_\phi$ and via the identification $H_\phi^0\simeq (Z^0_\phi)$ lift it to complete the basis $d_\phi^0$. If $c_\phi^0$ is another basis of $\alg$, the change of basis $\tau_\phi^0$ is now defined by
\be
\tau_\phi^0\equiv[h^0_\phi d_\phi^0/c_\phi^0],
\ee
and the relation between the Riemannian volume form on $G/T$, where $T\equiv\zeta(\phi)$ is the common stabilizer of non-singular flat connections $\phi$, and the form $\bigwedge(d_\phi^0)^*$ induced by $d_\phi^0$ now reads
\be
\bigwedge(d^0_\phi)^* = \f{\vol_{Z^0_\phi}(h^0_\phi)}{\tau^0_\phi}\ \f{\vol_{G/T}}{\vol_\alg(c^0_\phi)}.
\ee
Trivializing locally the bundle $G\rightarrow G/T$, we can as before pull the integral over the orbits $\cal{O}_\phi$ back to $G/T$:


\be
\calZ'(\Gamma,G) =\int_{\calF_{0}/G}\left(\int_{G/T}\vol_{G/T}\right)\f{1}{\tau_\phi^0\,\vol_\alg(c^0_\phi)}\ \f{\tau^1_\phi}{\tau^2_\phi}\ \f{\vol_{Z^0_\phi}(h^0_\phi)\vol_{T_\phi G^E}(c^1_\phi)\,\vol_{(B^2_\phi)^\perp}(h^2_\phi)}{\vol_{\alg^F}(c^2_\phi)}\,\bigwedge(h^1_\phi)^*.
\ee
The definition of the torsion is unchanged, $\tor_\phi \equiv\f{\tau^1_\phi}{\tau^0_\phi\tau^2_\phi}$, and the integral over the homogenous space $G/T$ gives a trivial factor $\Vol(G/T)$
\be
\calZ'(\Gamma,G) =\Vol(G/T)\ \bigl(\vol_\alg(l)\bigr)^{-\chi(\Gamma)} \int_{\calF_{0}/G} \tor_\phi\ \vol_{Z^0_\phi}(h_\phi^0)\,\vol_{(B^2_\phi)^\perp}(h^2_\phi)\ \bigwedge(h^1_\phi)^*,
\ee
where as before we used bases $c_\phi^k$ constructed from a fixed basis $l$ of the Lie algebra $\alg$. Choosing moreover $h_\phi^0$ and $h_\phi^2$ orthonormal, this simplifies to
\be\label{finalreducible}
\calZ'(\Gamma,G) =\Vol(G/T)\ \int_{\calF_{0}/G} \tor_\phi\ \bigwedge(h^1_\phi)^*,
\ee
This last equation is the generalization of (\ref{finalirreducible}) to the reducible case.
\bigskip

To summarize the results of this section, we have showed that
\begin{theorem}
If all non-singular $G$-connections $\phi\in\calF_{0}$ on the cell $2$-complex $\Gamma$ have the same isotropy type, and if the Reidemeister torsion is integrable on $\calF_{0}$, then the dominant part of the partition function of the flat spinfoam model is the volume of the moduli space of non-singular flat connections $\calF_{0}/G$ with respect to the Reidemeister torsion volume form, in the sense of (\ref{finalirreducible}) and (\ref{finalreducible}). 
\end{theorem}
\section{Conclusion}\label{sec:conclusion}

With a caveat related to the singularities of the representation variety of a finitely presented goup, we have described the structure of the `bubble divergences' arising in the flat spinfoam model: their divergence degree is given by the second Betti number in the natural twisted cohomology. This result has allowed us to identify a dominant part in the amplitudes, which is be given by the volume of the character variety of the fundamental group of the foam $\Gamma$, with a volume form given by the Reidemeister torsion of $\Gamma$.

One could try to use these results to define properly the
Ponzano-Regge model of Riemannian gravity on 3-manifolds, and its
higher dimensional analogues, as finite topological invariants. To this
aim, and also to settle the issue of singularities, it will be necessary to
focus attention to a specific class of foams $\Gamma$, such as special polyhedra. One could then attempt to relate the divergence degree $\Omega(\Gamma,G)$, and the dominant part $\calZ'(\Gamma,G)$, to the topology of the manifold itself. What is more, this would also permit a sharper comparison with topological quantum field theory, where similar divergences arise as a consequence of the continuum gauge symmetry \cite{witten-3d}. 

From the perspective of four-dimensional quantum gravity, we hope that our results will provide a useful basis for the study of the divergences of `non-flat' spinfoam models such as \cite{epr-long,new-model-fk}. Whether this hope is legitimate, the future will tell.

\appendix
\section*{Appendix: On non-integrable singularities}
\settocdepth{section}
\subsection*{Example of a non-integrable singularity}

The flat spinfoam model is defined by the choice of a finitely presented group $\pi$ and a structure group $G$. Consider $G=\SU(2)$ and \beq
\pi = \langle a,b,h\,\vert\,[a,h] = [b,h] =\unit \rangle.
\ee
It turns out that the divergence on singular connections of the Gaussian determinant in (\ref{gaussian int})  is \emph{not} integrable in this case. In other words, $(\pi,\SU(2))$ is an example which \emph{cannot} be handled by the method described in this paper. 
\medskip

The set of flat connections on the corresponding foam $\Gamma$ with three edges and two faces is determined by the relations
\beq
\calF = \left\{ (a,b,h)\in\SU(2)^3,\ [a,h] = [b,h] =\unit \right\}.
\ee
This set has two irreducible components. 

\begin{itemize}
\item
If $h$ is in the center of $\SU(2)$, i.e. $h=\pm \unit$, Then, $a,b$ can be arbitrary:
\beq
\calF_{\operatorname{irred}} \equiv \left\{ (a,b,\pm\unit),\ (a,b)\in\SU(2)^2\right\}.
\ee
These are the irreducible representations of $\pi=\pi_1(\Gamma)$ into $\SU(2)$. 
\item
If $h$ is not in the center, then $a,b$ and $h$ have to lie in a common $\U(1)$ subgroup of $\SU(2)$. These are the Abelian representations. If we write a generic element $g=\exp(i\psi \hat{n}\cdot\vec{\sigma})$, with $\hat{n}\in S^2$ the direction of the rotation and $\psi\in [0,\pi)$ its class angle, then
\beq
\calF_{\operatorname{red}} \equiv \left\{ (a,b,h)\in\SU(2)^3,\ a=\exp(\pm i\psi_a \hat{n}\cdot\vec{\sigma}), b=\exp(\pm i\psi_b \hat{n}\cdot\vec{\sigma}), h=\exp(i\psi_h \hat{n}\cdot\vec{\sigma})\right\}.
\ee
\end{itemize}
Quite obviously, $\calF_{\operatorname{irred}}$ is of dimension $6$, while $\calF_{\operatorname{red}}$ is $5$-dimensional. In both cases, these are the dimensions of the kernel of $\delta^1_\phi$ on non-singular flat connections. It follows that 
\beq
b^2_{\operatorname{irred}}(\Gamma,\SU(2)) = 3, \qquad\text{and}\qquad b^2_{\operatorname{red}}(\Gamma,\SU(2)) = 2.
\ee
So, we may expect the divergence to be controlled by non-singular irreducible flat connections, with $\Omega(\Gamma,\SU(2)) = 3$. However, this can be true only if the regular-indepedent integrals in \eqref{amplitude scaling} are finite. Here, we focus on the integral over the reducible representations and show that it is divergent.

\medskip

Choose as coordinates on $\calF_{\operatorname{red}}$ the variables $(\psi_a,\psi_b,\psi_h,\hat{n}_h)$, with $\hat{n}_h$ parametrized by spherical angles $(\theta_h,\varphi_h)$. Geometrically, we will perform a saddle point approximation corresponding to the fact that for a fixed direction $\hat{n}_h$, the integrals over $\hat{n}_a, \hat{n}_b$ are sharply peaked around $\pm\hat{n}_h$. The variables flowing away from the space of flat connections are the corresponding tangent vectors: $y_a = y^{\theta_a}\pp_{\theta_a}+y^{\varphi_a}\pp_{\varphi_a}$ and $y_b = y^{\theta_b}\pp_{\theta_b}+y^{\varphi_b}\pp_{\varphi_b}$. So we can write
\begin{align}
\calZ_\tau(\Gamma,\SU(2)) &= \int_{\SU(2)^3} da db dh\ K_\tau([a,h])\,K_\tau([b,h]),\\
& \begin{aligned}
\u{\tau\rightarrow 0}{\sim} 4\Lambda_\tau^6 \int \left(\int \exp\left\{-\f{\Vert \delta^1_\phi(y_a,y_b)\Vert^2}{4\tau}\right\}\,dy^{\theta_a} dy^{\varphi_a} dy^{\theta_b} dy^{\varphi_b}\right)\ &\sin^3\theta_h\,d\theta_h d\varphi_h \\
\times\ \sin^2\psi_a\,\sin^2\psi_b\,\sin^2\psi_h\,d\psi_a d\psi_b &d\psi_h + \calZ_{\tau}(\Gamma,\SU(2))_{\operatorname{irred}}, \end{aligned}
\end{align}
where a factor $\sin^2\theta_h$ comes from the evaluation of $(\sin\theta_a\, \sin\theta_b)$ of the Haar measure on the $\{+\hat{n}_a=+\hat{n}_b=\hat{n}_h\}$ component of $\calF$, and the factor $4$ accounts for the other components. The action of $\delta^1_\phi$ on $y_a$ and $y_b$ is similar to that discussed in the 2-torus example \eqref{delta1torus}. The determinant of the quadratic form $\Vert\delta^1_\phi\Vert^2$ in the basis $b_\phi \equiv (\pp_{\theta_a}, \pp_{\varphi_a}, \pp_{\theta_b}, \pp_{\varphi_b})$ can be evaluated with the same tools, yielding 
\beq
\sqrt{\det(\Vert\delta^1_\phi\Vert^2,b_\phi)} = 16\,\bigl(\sin\theta_h\,\sin^2\psi_a\,\sin^2\psi_h\bigr)\,\bigl(\sin\theta_h\,\sin^2\psi_b\,\sin^2\psi_h\bigr).
\ee
Thus, our method produces the following tentative asymptotic equivalent for $\calZ_{\tau}(\Gamma,\SU(2))_{\operatorname{red}}$:
\beq
\f 14\Lambda_\tau^2\ \left(\int_{S^2}\vol_{S^2}\right)\ \int_{[0,\pi]^3} \f{d\psi_a\ d\psi_b\ d\psi_h}{\sin^2\psi_h}.
\ee
The integral over $S^2$ corresponds to the integral over the orbits of gauge transformations, which here correspond to rotations of the common direction $\hat{n}_h$ of the three group elements, the class angles remaining fixed. In other words, the moduli space in this case is parametrized by the class angles $\psi_a, \psi_b$ and $\psi_h$. However, one can see that the remaining integral, over the moduli space, is \emph{divergent}. This means that the correct equivalent of the partition function as $\tau$ goes to zero is \emph{not} this one, and that the scaling is \emph{not} given by $\Lambda_\tau^{b^2_0(\Gamma,G)}$ (nor by any monomial in $\Lambda_\tau$) in this pathological case.


\subsection*{A non-integrable singularity in a Laplace integral}

Let us conclude this appendix by a very simple example showing how a non-integrable singularity can spoil the Laplace estimate for saddle point integrals, borrowed from \cite{cassanas}. Consider the numerical integal

\be
z_\tau\equiv\int_{\mathbb{R}^2}dxdy\ e^{-\f{(xy)^2}{\tau}}.
\ee
Here, the critical set is the `cross' $\{x=0\}\cup\{y=0\}$, and has a singularity at $(x,y)=(0,0)$. A naive application of the Laplace approximation would give $z_\tau\propto\tau^{1/2}$ as $\tau\rightarrow0$, since the orthogonal space to the critical set is one-dimensional. However, this is not the correct estimate, which turns out to be $z_\tau\propto\tau^{1/2}\ln\tau$. In this case, indeed, integrating along the `normal fibers' yields a non-integrable singularity: formally,
\be
z_\tau=\int_{\mathbb{R}}dx\left(\int_{\mathbb{R}}dy\ e^{-\f{(xy)^2}{\tau}}\right)=\sqrt{\pi\tau}\int_{\mathbb{R}}\f{dx}{\vert x\vert}.
\ee
This behaviour is stricly analogous to the case where the Reidemeister torsion in not integrable in the neighborhood of singular connections: it signals the breakdown of the Laplace approximation, and calls for a more sophisticated analysis. 

\begin{thebibliography}{10}

\bibitem{rovelli-book}
C.~Rovelli, {\em Quantum Gravity}.
\newblock Cambridge University Press, 2004.

\bibitem{thiemann-book}
T.~Thiemann, {\em {Modern canonical quantum general relativity}}.
\newblock Cambridge University Press, 2007, [arXiv:gr-qc/0110034].

\bibitem{baez-bf-spinfoam}
J.~C. Baez, ``{An introduction to spin foam models of BF theory and quantum
  gravity},'' {\em Lect. Notes Phys.}, vol.~543, pp.~25--94, 2000,
  gr-qc/9905087.

\bibitem{perez-spin-foam-representation}
A.~Perez, ``{The spin-foam-representation of loop quantum gravity},'' 2006, [arXiv:gr-qc/0601095].

\bibitem{fk-action-principle}
L.~Freidel and K.~Krasnov, ``{Spin foam models and the classical action
  principle},'' {\em Adv. Theor. Math. Phys.}, vol.~2, pp.~1183--1247, 1999,
  [arXiv:hep-th/9807092].

\bibitem{plebanski}
J.~F. Plebanski, ``{On the separation of Einsteinian substructures},'' {\em J.
  Math. Phys.}, vol.~18, pp.~2511--2520, 1977.

\bibitem{epr-long}
J.~Engle, R.~Pereira, and C.~Rovelli, ``{Flipped spinfoam vertex and loop
  gravity},'' {\em Nucl. Phys.}, vol.~B798, pp.~251--290, 2008, [arXiv:0708.1236].

\bibitem{new-model-fk}
L.~Freidel and K.~Krasnov, ``{A New Spin Foam Model for 4d Gravity},'' {\em
  Class. Quant. Grav.}, vol.~25, p.~125018, 2008, [arXiv:0708.1595].

\bibitem{PR}
G.~Ponzano and T.~Regge, ``{Semi-classical limit of Racah coefficients},'' {\em
  Spectroscopic and group theoretical methods in physics (F. Bloch, ed.),
  North-Holland, Amsterdam}, 1968.

\bibitem{perez-BC-bubble}
A.~Perez and C.~Rovelli, ``{A spin foam model without bubble divergences},''
  {\em Nucl. Phys.}, vol.~B599, pp.~255--282, 2001, [arXiv:gr-qc/0006107].

\bibitem{freidel-louapre-diffeo}
L.~Freidel and D.~Louapre, ``{Diffeomorphisms and spin foam models},'' {\em
  Nucl. Phys. B}, vol.~662, no.~1-2, pp.~279--298, 2003.

\bibitem{freidel-gurau-oriti}
L.~Freidel, R.~Gurau, and D.~Oriti, ``{Group field theory renormalization - the
  3d case: power counting of divergences},'' {\em Phys. Rev.}, vol.~D80,
  p.~044007, 2009, [arXiv:0905.3772].

\bibitem{ben-geloun-abelien}
J.~Ben~Geloun, T.~Krajewski, J.~Magnen, and V.~Rivasseau, ``{Linearized Group
  Field Theory and Power Counting Theorems},'' 2010, [arXiv:1002.3592].

\bibitem{cell-homology}
V.~Bonzom and M.~Smerlak, ``{Bubble divergences from cellular homology},''
  2010, [arXiv:1004.5196].

\bibitem{witten-2d-YM}
E.~Witten, ``{On quantum gauge theories in two dimensions},'' {\em Commun.
  Math. Phys.}, vol.~141, no.~1, pp.~153--209, 1991.

\bibitem{goldman}
W.~Goldman, ``The symplectic nature of fundamental groups,'' {\em Adv. Math.},
  vol.~54, p.~200, 1984.

\bibitem{turaev-viro}
V.~G. Turaev and O.~Y. Viro, ``{State sum invariants of 3 manifolds and quantum
  6j symbols},'' {\em Topology}, vol.~31, pp.~865--902, 1992.

\bibitem{crane-kauffman-yetter}
L.~Crane, L.~H. Kauffman, and D.~Yetter, ``{Evaluating the Crane-Yetter
  invariant},'' 1993, [arXiv:hep-th/9309063].

\bibitem{boulatov-model}
D.~V. Boulatov, ``{A Model of three-dimensional lattice gravity},'' {\em Mod.
  Phys. Lett.}, vol.~A7, pp.~1629--1646, 1992, [arXiv:hep-th/9202074].

\bibitem{ooguri4d}
H.~Ooguri, ``{Topological lattice models in four-dimensions},'' {\em Mod. Phys.
  Lett.}, vol.~A7, pp.~2799--2810, 1992, [arXiv:hep-th/9205090].

\bibitem{freidel-gft}
L.~Freidel, ``{Group field theory: An overview},'' {\em Int. J. Theor. Phys.},
  vol.~44, pp.~1769--1783, 2005, [arXiv:hep-th/0505016].

\bibitem{oriti-gft-review-06}
D.~Oriti, ``{The group field theory approach to quantum gravity},'' 2006,
  [arXiv:gr-qc/0607032].

\bibitem{barrett-PR}
J.~W. Barrett and I.~Naish-Guzman, ``{The Ponzano-Regge model},'' {\em Class.
  Quant. Grav.}, vol.~26, p.~155014, 2009, [arXiv:0803.3319].

\bibitem{blau-thompson-bf}
M.~Blau and G.~Thompson, ``Topological gauge theories of antisymmetric tensor
  fields,'' {\em Annals Phys.}, vol.~205, pp.~130--172, 1991.

\bibitem{gurau-colored-gft}
R.~Gurau, ``{Colored Group Field Theory},'' 2009, [arXiv:0907.2582].


\bibitem{Forman-small}
R.~Forman, ``{Small volume limits of 2-d Yang-Mills},'' {\em Commun. Math.
  Phys.}, vol.~151, pp.~39--52, 1993.

\bibitem{atiyah-bott-2d-YM}
M.~F. Atiyah and R.~Bott, ``{The Yang-Mills equations over Riemann surfaces},''
  {\em Phil. Trans. Roy. Soc. Lond.}, vol.~A308, pp.~523--615, 1982.

\bibitem{sengupta-singularities-2d}
A.~Sengupta, ``{The volume measure for flat connections as limit of the
  Yang-Mills measure},'' {\em J. Geom. Phys.}, vol.~47, no.~4, pp.~398--426,
  2003.
  
\bibitem{witten-3d}
E. Witten, ``{Topology-changing amplitudes in (2+1)-dimensional gravity}", \emph{Nuclear Phys. B}, vol. 323, pp. 113--140, 1989.

\bibitem{gegenberg-partition-function-bf}
J.~Gegenberg and G.~Kunstatter, ``{The Partition function for topological field
  theories},'' {\em Ann. Phys.}, vol.~231, pp.~270--289, 1994, [arXiv:hep-th/9304016].

\bibitem{dubois private}
J. Dubois, {\em private communication}. 

\bibitem{turaev-book}
V.~Turaev, {\em {Introduction to combinatorial torsions}},
\newblock Birkhauser, 2001.

\bibitem{cassanas}
R.~Cassanas, {\em unpublished notes}.

\end{thebibliography}

\end{document}